\documentclass [12pt]{article}
\usepackage{graphics}   
\usepackage[dvips]{graphicx}
\usepackage{amssymb,amsfonts,latexsym,amsmath,amsthm,times}
\usepackage[left=1.1in, right=0.9in, top=1in, bottom=0.9in, headheight=13.6pt]{geometry}
\usepackage{epsfig}
\usepackage{fancyhdr}
\usepackage{color}
\usepackage[english]{babel}
\usepackage[dvips]{graphicx}
\usepackage{subfigure}  

\newcommand{\Real}{\mathbb{R}}
\newcommand{\Natural}{\mathbb{N}}
\newcommand{\Numbers}{\mathbb{Z}}

\newcommand {\phivec}{\phi}
\newcommand {\bfx}{x}

\newcommand {\bfz}{z}
\newcommand {\varphivec}{{\boldsymbol{\varphi}}}
\newcommand {\thetavec}{{\boldsymbol{\theta}}}
\newcommand {\lambdavec}{{\boldsymbol{\lambda}}}
\newcommand{\pd}{{\partial}}
\newcommand {\col}{{\mathrm{col}}}
\newcommand{\dist}{\mathrm{dist}}
\newcommand{\norm}[1]{\left\Vert#1\right\Vert}

\numberwithin{equation}{section}

\newtheorem{theorem}{Theorem}

\newtheorem{corollary}[theorem]{Corollary}

\newtheorem{definition}[theorem]{Definition}

\newtheorem{proposition}[theorem]{Proposition}
\newtheorem{remark}[theorem]{Remark}

\begin{document}

\footnotesize {\flushleft \mbox{\bf \textit{Math. Model. Nat.
Phenom.}}}
 \\
\mbox{\textit{{\bf Vol. 5, No. 3, 2010, pp. 146-184}}}

\thispagestyle{plain}

\vspace*{2cm} \normalsize \centerline{\Large \bf Observers for Canonic Models of Neural Oscillators}

\vspace*{1cm}

\centerline{\bf D. Fairhurst$^a$, I. Tyukin $^{a,c,d}$\footnote{Corresponding
author. E-mail: I.Tyukin@le.ac.uk}, H. Nijmeijer$^b$, and C. van Leeuwen$^c$}

\vspace*{0.5cm}

\centerline{$^a$ Department of Mathematics,
 University of Leicester, University Road, LE1 7RH, UK}

\centerline{$^b$ Department of Mechanical Engineering, Eindhoven University of Technology,}
 \centerline{ P.O. Box 513 5600 MB,  Eindhoven, The Netherlands}

\centerline{$^c$ RIKEN (Institute for Physical and Chemical Research) Brain Science Institute,}
\centerline{ 2-1, Hirosawa, Wako-shi, Saitama, 351-0198, Japan}

\centerline{$^d$ Deptartment of Automation and Control Processes, St-Petersburg State University}
 \centerline{of Electrical Engineering, Prof. Popova str. 5, 197376, Russia}


\vspace*{1cm}

\noindent {\bf Abstract.}
We consider the problem of state and parameter estimation for a
class of nonlinear oscillators defined as a system of coupled
nonlinear ordinary differential equations. Observable variables
are limited to a few components of state vector and an input
signal. This class of systems describes a set of canonic models
governing the dynamics of evoked potential in neural membranes,
including Hodgkin-Huxley, Hindmarsh-Rose, FitzHugh-Nagumo, and
Morris-Lecar models. We consider the problem of state and
parameter reconstruction for these models within the classical
framework of observer design. This framework offers
computationally-efficient solutions to the problem of state and
parameter reconstruction of
 a system of nonlinear differential equations, provided that these
 equations are in the so-called adaptive observer canonic form.
We show that despite typical neural oscillators being locally
observable they are not in the adaptive canonic
observer form. Furthermore, we show that no
parameter-independent diffeomorphism exists such that the
original equations of these models can be transformed into the
adaptive canonic observer form. We demonstrate,
 however, that for the class of Hindmarsh-Rose and FitzHugh-Nagumo
models, parameter-dependent coordinate transformations can be
used to render these systems into the adaptive observer
canonical form. This allows reconstruction, at least partially
and up to a (bi)linear transformation, of unknown state and
parameter values with exponential rate of convergence. In order
to avoid the problem of only partial reconstruction and at the
same time to be able to deal with more general nonlinear models
in which the unknown parameters enter the system nonlinearly,
we present a new method for state and parameter
 reconstruction for these systems. The method combines
advantages of standard Lyapunov-based design with more flexible
design and analysis techniques based on the notions of positive
invariance and small-gain theorems. We show that this
flexibility allows to overcome ill-conditioning and
non-uniqueness issues arising in this problem. Effectiveness of
our method is illustrated with simple numerical examples.

\vspace*{0.5cm}




\vspace*{1cm}

\setcounter{equation}{0}
\section*{Notations and Nomenclature}

The following notational conventions are used throughout the
paper:
\begin{itemize}
\item $\Real$ is the field of real numbers.

\item $\Real_{>0} = \{x\in\Real\ |\ x>0\}$.

\item $\Numbers$ denotes the set of integers, and
    $\Natural$ stands for the set of positive integers.

\item The Euclidian norm of $x \in \Real^n$ is denoted by
    $\norm{x}$.

\item $\mathcal{C}^{r}$ denotes the space of continuous
    functions that are at least $r$ times differentiable.

\item Let  $h:\Real^n\rightarrow\Real$ be a differentiable
    function and $f:\Real^n\rightarrow\Real^n$. Then
    $L_{f}h(x)$, or simply $L_f h$, is the Lie derivative
    of $h$ with respect to $f$:
    \[
    L_f h(x)=\frac{\pd h}{\pd x} f(x)
    \]

\item Let  $f, g :\Real^n\rightarrow\Real^n$ be
    differentiable vector-fields. Then the symbol $[f,g]$
    stands for the Lie bracket:
    \[
    [f,g]=\frac{\pd f}{\pd x} g - \frac{\pd g}{\pd x} f
    \]
 The adjoint representation of the Lie bracket is defined
 as
    \[
    \mathrm{ad}_f^0 g = g, \ \mathrm{ad}_f^k g = [f,\mathrm{ad}_f^{k-1} g]
    \]

\item Let $\mathcal{A}$ be a subset of $\Real^n$, then for
    all $x\in\Real^n$, we define
    $\mathrm{dist}(\mathcal{A},x)=\inf_{q\in\mathcal{A}}\|x-q\|$.

\item $\mathcal{O}(\cdot)$ denotes a function such that
    $\lim_{s\rightarrow 0} \mathcal{O}(s)/s=d$,
    $d\in\Real$, $d\neq 0$.

\item Finally, let $\epsilon \in \Real_{>0}$, then
    $\norm{\bfx}_\epsilon$ stands for the following:
\[
\norm{\bfx}_\epsilon = \left\{
\begin{array}{ll}
\norm{\bfx}-\epsilon, & \norm{\bfx} > \epsilon,\\
0,& \norm{\bfx} \leq \epsilon.
\end{array}\right.
\]
\end{itemize}

\vspace*{0.5cm}
\section{Introduction}

Mathematical modelling of brain processes and function is recognized as an important
tool of modern neuroscience \cite{PNAS:Izhikevich:2008}. It
allows us to predict, analyze and understand intricate
processes of neural computations, without invoking technically
involving and costly experiments. Successful examples include
but are not limited to modelling the memory function
\cite{Biosystems:Borisyuk:2006}, \cite{NC:Borisyuk:2006} and
the mechanisms of phase-resetting in olivo-cerebellar networks
\cite{Kazantsev:2004}. Availability of quantitatively accurate
models of individual neural cells is an important prerequisite of
such studies.

The majority of available models of individual biological
neurons are the systems of ordinary differential equations
describing the cell's response to stimulation; their parameters
characterize variables such as time constants, conductances,
and response thresholds,  important for relating the model
responses to behavior of biological cells. Typically two
general classes of models co-exist: phenomenological and
mathematical ones. Models of the first class, such as e.g. the
Hodgkin-Huxley equations, claim biological plausibility,
whereas models of the second class are more abstract
mathematical reductions without explicit relation of all of
their variables to physical quantities such as conductances and
ionic currents (see Table \ref{tab:neuron_models}).
\begin{table}[b!]
\normalsize
  \centering \vskip -7mm
  \caption{Examples of typical mathematical models of spiking single neurons. Biologically plausible equations of membrane potential generation in a giant axon of a squid (left panel), and a reduction of these equations to an oscillator with polynomial right-hand side (right panel)}\label{tab:neuron_models}
\vskip 1mm {\small
  \begin{tabular*}{\textwidth}{@{\extracolsep{\fill}}|c|c|}
    \hline
    & \\
     Hodgkin-Huxley Model \cite{Hodgkin_Huxley} &  Hindmarsh-Rose Model \cite{Hindmarsh_and_Rose} \\
     & \\
    \hline
    \begin{minipage}[l]{0.65\linewidth}
    \vskip -2mm \normalsize
    \begin{eqnarray}\label{eq:HH}
    \dot{v}&=& I(t) - \left[\theta_1 m^3 h(v+\theta_2)+\theta_3 n^4(v+\theta_4)+ \theta_5 v + \theta_6 \right]\nonumber \\
    \dot{m}&=& (1-m) \varphi\left(\frac{v+\theta_7}{\theta_8}\right)-m \ \theta_9 \exp\frac{v}{\theta_{10}} \nonumber \\
    \dot{n}&=& (1-n)\theta_{11} \varphi \left(\frac{v+\theta_{12}}{\theta_{13}}\right) -n \ \theta_{14}\exp\frac{v}{\theta_{15}} \\
    \dot{h}&=&(1-h)\theta_{16}\exp\frac{v}{\theta_{17}} - h/\left(1+\exp\frac{v+\theta_{18}}{\theta_{19}}\right)\nonumber
     \end{eqnarray}
    $\varphi(x)={x}/({\exp{x}-1})$, $\theta_1,...,\theta_{19}$ -- parameters, $I:\Real\rightarrow\Real$ --  input current
    \end{minipage} &
    \begin{minipage}[l]{0.29\linewidth}
    \vskip -2mm \normalsize
    \begin{eqnarray}\label{eq:HR}
    \dot{v}&=& \theta_1 (\theta_2 r - f(v)) + I(t)\nonumber \\
    \dot{r}&=& \theta_3 (g(v) - \theta_4 r)
    \end{eqnarray}
    $f(v)$ and $g(v)$ are polynomials:
    \begin{eqnarray}
    f(v)&=&\theta_5 +\theta_6 v + \nonumber \\
    &=&\theta_7 v^2 + \theta_8 v^3; \nonumber \\
    g(v)&=&\theta_{9} + \theta_{10} v + \theta_{11} v^2\nonumber
    \end{eqnarray}
   $I:\Real\rightarrow\Real$ is the external input, $\theta_1,...,\theta_{11}$ -- parameters
    \end{minipage}
        \\
        & \\
 \hline
  \end{tabular*}
} \vspace{-4mm}
\end{table}

Despite these
differences, these models admit a common general
description which will be referred to as {\it canonic}.
 In particular, the dynamics of a typical neuron are governed by the following set of equations
\begin{equation}\label{eq:non_canonic}
\begin{split}
\dot{v}&= {\sum}_j \ \varphi_j(v,t) p_j(r) \theta_{j}+I(t);\\
\dot{r}_i&= - a_i(v,\theta,t) r_i + b_i(v,\theta,t); \\
\theta&=(\theta_1,\theta_2,\dots).
\end{split}
\end{equation}
in which the variable $v$ is the membrane potential, and $r_i$
are the gating variables of which the values are not available
for direct observation. Functions $\varphi_j(\cdot,\cdot)$,
$p_j(\cdot)\in\mathcal{C}^{1}$  are assumed to be known; they
model components of specific ionic conductances. Functions
$a_i(\cdot)$, $b_i(\cdot)\in\mathcal{C}^{1}$ are also known,
yet they depend on the unknown parameter vector $\theta$.
System (\ref{eq:non_canonic}) is a typical conductance-based
description of the evoked potential generation in neural
membranes \cite{Izhikevich:2007}. It is also an obvious
generalization of many purely mathematical models of spike
generation such as the FitzHugh-Nagumo \cite{FitzHugh} or the
Hindmarsh-Rose equations \cite{Hindmarsh_and_Rose}. In this
sense systems (\ref{eq:non_canonic}) represent typical building
blocks in the modelling literature.

In order to be able to model the behavior of large numbers of
individual cells of which the input-output responses are
described by (\ref{eq:non_canonic}), computational tools for
automated fitting of models of neurons to data are needed.
These tools are the algorithms for state and parameter
reconstruction of (\ref{eq:non_canonic}) from the available
measurements of $v(t)$ and $I(t)$ over time.

Fitting parameters of nonlinear ordinary differential equations
to data is recognized as a hard computational problem
\cite{Brewer:2008} that ``has not been yet treated in full
generality" \cite{Ljung:2008}. Within the field of
neuroscience,  conventional methods for fitting parameters of
model neurons to measured data are often restricted to
hand-tuning or exhaustive trial-and-error search in the space
of model parameters \cite{Prinz:2003}. Even though these strategies
allow careful and detailed exploration in the space of
parameters they suffer from the same problem -- the curse of
dimensionality.

Available alternatives, recognizing obvious nonlinearity of the
original problem, propose to reformulate the original estimation problem as that of searching for the parameters of a system of difference equations approximating solutions of (\ref{eq:non_canonic}) \cite{SIAM:Abarbanel}; or predominantly offer search-based optimization heuristics (see \cite{van_geit:2008}
 for a detailed review) as the main tool for automated fitting of neural models.
 Straightforward exhaustive-search approaches however are limited to varying only few
 model parameters over sparse grids, e.g. as in \cite{Prinz:2003} where $8$ parameters
were split into $6$ bands. Coarseness of this parametrization
leads to non-uniqueness of signal representation, leaving room
for uncertainty and inability to distinguish
 between subtle changes in the cell. More fine-grained search algorithms are currently
infeasible, technically  speaking. Other heuristics, such as
evolutionary algorithms,
 are examined in \cite{PLOS_Bio:Achard:2006}. According to \cite{PLOS_Bio:Achard:2006},
replacing exhaustive search with evolutionary algorithms allows
to increase the number of varying parameters to $24$. Yet,
computational complexity of the problem still delimits the
search to sparse grids ($6$ bands per single parameter) and
requires days of simulation by a cluster of $10$ Apple 2.3 GHz
nodes.  Furthermore, because all these strategies are heuristic,
accuracy of final results is not guaranteed.

The main aim of this article is to present a feasible
substitute to these heuristic strategies for automatic
reconstruction of state and parameters of canonic neural models
(\ref{eq:non_canonic}). To develop computationally efficient
procedures for  state and parameter reconstruction of
(\ref{eq:non_canonic}) we propose to exploit the wealth of
system-identification and estimation approaches from the domain
of control theory. These approaches are based on the
system-theoretic concepts of observability and identifiability
\cite{Nijmeijer_90}, \cite{Isidory_99},\cite{Ljung_99} from
control theory, and the notions of Lyapunov stability
\cite{Lyapunov} and weakly attracting sets \cite{Milnor_1985}.
The advantage of using these approaches is that there is an
abundance of algorithms (observers) already developed within
the domain of control. These algorithms guarantee asymptotic
and stable reconstruction of unmeasured quantities from the
available observations, provided that the system equations are
in an {\it adaptive observer canonical form}. Moreover, this
reconstruction can be made exponentially fast without the need
of substantial computational recourses. We study if system
(\ref{eq:non_canonic}) is at all observable with respect to the
output $v$, that is if its state and parameters can be
reconstructed from observations of $v$. We present and analyze
typical algorithms (adaptive observers) that are available in
the literature. We show that for a large class of mathematical
models of neural oscillators at least a part of the model
parameters can be reconstructed exponentially fast.

In order to deal with more general classes of models and also
to recover the rest of the model parameters we introduce a
novel observer scheme. This scheme benefits from 1) the
efficiency of uniformly converging estimation procedures
(stable observers), 2) success of explorative search strategies
in global optimization by allowing unstable convergence along
dense trajectories, and 3) the power of qualitative analysis of
dynamical systems. We present a general description of this
observer and list its asymptotic properties. The theory of this
new class of algorithms is based on the results of our previous
studies in the domain of unstable convergence
\cite{SIAM_non_uniform_attractivity},
\cite{Adaptive_Observers}. We will present examples to
demonstrate the performance of these algorithms.

The paper is organized as follows. In Section
2 we provide the basic notions of
observability from the domain of mathematical control, test if
typical canonical neural oscillators are observable, and
present two major classes of systems (canonical forms) for
which computationally efficient reconstruction procedures are
available. In Section 3 we analyze
the applicability of standard observers to the problem of
reconstructing all unmeasured variables and parameters of
typical models of neurons. We present two special cases in
which such reconstruction is possible. In Section
4 we provide a description and asymptotic
properties of our algorithm that applies to the most general
subset of models (\ref{eq:non_canonic}). Section
\ref{sec:examples} contains examples of application of the
considered observers, and Section 5 concludes
the paper. Proofs of the main technical statements are
presented in the Appendix.


\vspace*{0.5cm}
\setcounter{equation}{0}

\section{Observer-based approaches to the problem of state and parameter estimation}\label{sec:observer}

Let us consider the following class of dynamical systems
\begin{equation}\label{eq:system:general}
\begin{split}
 \dot{x} &= f(x,\theta) + g(x,\theta)u(t), \quad x(t_0)\in\Omega_x\subset\Real^n \\
  y &= h(x), \quad x\in \mathbb{R}^n, \quad  \theta\in\Real^d, \quad y\in \mathbb{R}
\end{split}
\end{equation}
where $f,g:\Real^n\times\Real^m\rightarrow\Real^n$,
$h:\Real^n\rightarrow\Real$ are
smooth functions\footnote{Let us recall that a function is smooth in $G$ if for every $x\in G$ and $n\in N$ the function $\frac{d^n}{dx^n}f(x)$ is always defined.}, and $u:\Real\rightarrow\Real$. Variable $x$ stands
for the state vector, $u\in\mathcal{U}\subset
C^{1}[t_0,\infty)$ is the known input, $\theta\in\Real^m$ is the vector
of unknown parameters, and $y$ is the output of
(\ref{eq:system:general}).  System (\ref{eq:system:general})
includes equations (\ref{eq:non_canonic}) as a subclass and in
this respect can be considered as plausible generalizations.
Obviously, conclusions about (\ref{eq:system:general}) should
be valid for systems (\ref{eq:non_canonic}) as well.

Given that the right-hand side of (\ref{eq:system:general}) is
differentiable, for any $x'\in\Omega_x$,
$u\in\mathcal{C}^{1}[t_0,\infty)$ there exists a time interval
$\mathcal{T}=[t_0,t_1]$, $t_1>t_0$ such that a solution
$x(t,x')$ of (\ref{eq:system:general}) passing through $x'$ at
$t_0$ exists for all $t\in\mathcal{T}$. Hence $y(t)=h(x(t))$ is
defined for all $t\in\mathcal{T}$. For the sake of convenience
we will assume that the interval $\mathcal{T}$ of the solutions
is large enough or even coincides with $[t_0,\infty)$ when
necessary.

We are interested in finding an answer to the following
question: suppose that we are able to measure the values of
$y(t)$ and $u(t)$ precisely; wether and how the values of $x'$ and
parameter vector $\theta$ can be recovered from the
observations of $y(t)$ and $u(t)$ over a finite subinterval of
$\mathcal{T}$? A natural framework to answer to these questions
is offered by the concept of {\it observability} \cite{Nijmeijer_90}.
\vspace*{0.25cm}
\begin{definition}[Observability]
\label{defi:observability_general}
 Two states $x_1,x_2\in\Real^n$ are said to be indistinguishable (denoted by $x_1\mathcal{I}x_2$)
  for (\ref{eq:system:general}) if  for every admissible input function $u$ the
  output function $t\rightarrow y(t,0,x_1,u)$, $t\geq 0$ of the system for initial state $x(0)=x_1$,
   and the output function $t\rightarrow y(t,0,x_2,u)$, $t\geq 0$ of the system for initial state $x(0)=x_2$,
    are identical on their common domain of definition.
    The system is called observable if $x_1\mathcal{I}x_2$ implies $x_1=x_2$.
\end{definition}
According to Definition \ref{defi:observability_general},
observability of a dynamical system implies that the values of its
state, $x(t)$, $t\in[t_1,t_2]$ are completely determined by inputs
and outputs $u(t)$, $y(t)$ over $[t_1,t_2]$. Although this
definition does not account for any unknown parameter vectors, one
can easily see that the very same definition can be used for
parameterized systems as well. Indeed, extending original equations
(\ref{eq:system:general}) by including parameter  vector $\theta$ as
a component of the extended state vector $\tilde{x}=(x,\theta)^T$
results in
\begin{equation}\label{eq:system:general:1}
\begin{split}
 \dot{x} &= f(x,\theta) + g(x,\theta)u(t), \\
 \dot{\theta}&=0\\
  y &= h(x), \quad x(t_0)\in\Omega_x\subset\Real^n
\end{split}
\end{equation}
or, similarly, in
\begin{equation}\label{eq:system:general:2}
\begin{split}
 \dot{\tilde{x}} &= \tilde{f}(\tilde{x}) + \tilde{g}(\tilde{x}) u(t) \\
   y &= \tilde{h}(\tilde{x}), \quad  \tilde{x}(t_0)=(x(t_0),\theta)^{T}\in\Omega_{\tilde{x}}\subset\Real^{n+d}
\end{split}
\end{equation}
where  $\tilde{f}(\tilde{x})=(f(x,\theta),0)^{T}$,
$\tilde{g}(\tilde{x})=(g(x,\theta),0)^{T}$, and
$\tilde{h}(\tilde{x})=(h(x),0)$. All uncertainties in
(\ref{eq:system:general}), (\ref{eq:system:general:1}),
including the parameter vector $\theta$, are now combined into
the state vector of (\ref{eq:system:general:2}). Hence the
problem of state and parameter reconstruction of
(\ref{eq:system:general}) can be viewed as that of recovering
the values of state for (\ref{eq:system:general:2}).

Definition \ref{defi:observability_general} characterizes observability as a global property of a dynamical system. Sometimes, however, global observability of a system in $\Real^n$ is not necessarily needed. Instead of asking if {\it every} point in the system's state space is distinguishable from any other point it may be sufficient to know if the system's states are distinguishable in some neighborhood of a given point. This necessitates the notion of local observability  \cite{Nijmeijer_90}.

Let $V$ be an open subset of $\Real^n$.  Two states $x_1,x_2\in V$ are said to be indistinguishable (denoted by $x_1\mathcal{I^{V}}x_2$) on $V$
  for (\ref{eq:system:general}) if  for every admissible input function $u: [0,T]\rightarrow \Real$ with the property that the solutions $x(t,0,x_1,u)$, and $x(t,0,x_2,u)$ both remain in $V$ for $t\leq T$ the
  output function $t\rightarrow y(t,0,x_1,u)$, $t\geq 0$ of the system for initial state $x(0)=x_1$,
   and the output function $t\rightarrow y(t,0,x_2,u)$, $t\geq 0$ of the system for initial state $x(0)=x_2$,
    are identical for $0\leq t\leq T$ on their common domain of definition.

\begin{definition}[Local observability \cite{Nijmeijer_90}]
\label{defi:observability_local}
    The system is called locally observable at $x_0$ if there exists a neighborhood $W\subset \Real^n$ of $x_0$  such that for every neighborhood $V\subset W$ of $x_0$ the relation $x_0\mathcal{I^V}x_1$ implies $x_0=x_1$. The system is locally observable if it is observable at each $x_0$.
\end{definition}

A number of observability tests are available that, given the
functions ${h}$,  ${f}$  in the right-hand side of
(\ref{eq:system:general:2}), indicate if a given system is
observable. Particular formulations of these tests may vary depending on whether e.g. the functions $f,g$  are analytic or time-invariant (inputs are constants).

In this article we will restrict our  attention to those systems (\ref{eq:system:general:1}) in which the inputs $u(t)$ are constants. In this case we can replace the function $u(t)$ with an unknown parameter, and system (\ref{eq:system:general:1}) can be viewed as a system (\ref{eq:system:general:2}) yet without inputs. One of the most common observability tests for this class of autonomous systems is given below (see also \cite{Nijmeijer_90}, Theorem 3.32):
\vspace*{0.25cm}
\begin{proposition} [Observability test (Corollary 3.33, \cite{Nijmeijer_90})]
 \label{defi_observability} System (\ref{eq:system:general:2})
is locally observable at a point $x^o
\in U \subset \mathbb{R}^{n+d}$ if
\begin{eqnarray}\label{eq:observability_condition}
 {\rm rank}
 \frac{\partial}{\partial \tilde{x}}
 \left(
   \tilde{h}(\tilde{x})\quad L_{\tilde{f}} \tilde{h}(\tilde{x})\quad L^2_{\tilde{f}} \tilde{h}(\tilde{x})\quad ...\quad L^{n+d-1}_{\tilde{f}}\tilde{h}(\tilde{x})
 \right)^T
 = n+d, \quad \forall \ \tilde{x} \in U
\end{eqnarray}
\end{proposition}

In what follows we shall use the test above to determine if the models of neural dynamics are at all observable.

\subsection{Local observability of neural oscillators}

We start our observability analysis by applying the local
observability test (\ref{eq:observability_condition}) to the
Hindmarsh-Rose model (\ref{eq:HR}). In order to do so we shall
extend the system state space so that unknown parameters are
the components of the extended state vector.
 In the case of the Hindmarsh-Rose model this procedure leads to the following extended system of equations:
\begin{eqnarray}\label{eqn_hindmarsh_rose}
 \left(
  \begin{array}{c}
   \dot{x}_1
  \\
   \dot{x}_2
  \\
   \dot{\theta}_{13}
  \\
   \dot{\theta}_{12}
  \\
   \dot{\theta}_{11}
  \\
   \dot{\theta}_{10}
  \\
   \dot{\theta}_{22}
  \\
   \dot{\theta}_{21}
  \\
   \dot{\lambda}
  \end{array}
  \right)
 =
 \left(
  \begin{array}{c}
   \theta_{13} x_1^3 + \theta_{12} x_1^2 + \theta_{11} x_1 + \theta_{10} + x_2
  \\
   -\lambda x_2 + \theta_{22} x_1^2 + \theta_{21} x_1
  \\
   0
  \\
   0
  \\
   0
  \\
   0
  \\
   0
  \\
   0
  \\
   0
  \end{array}
  \right)
\end{eqnarray}

\begin{figure}
 \centering
 \subfigure[$\delta = 10^{20}$]{\resizebox{5cm}{5cm}{\includegraphics{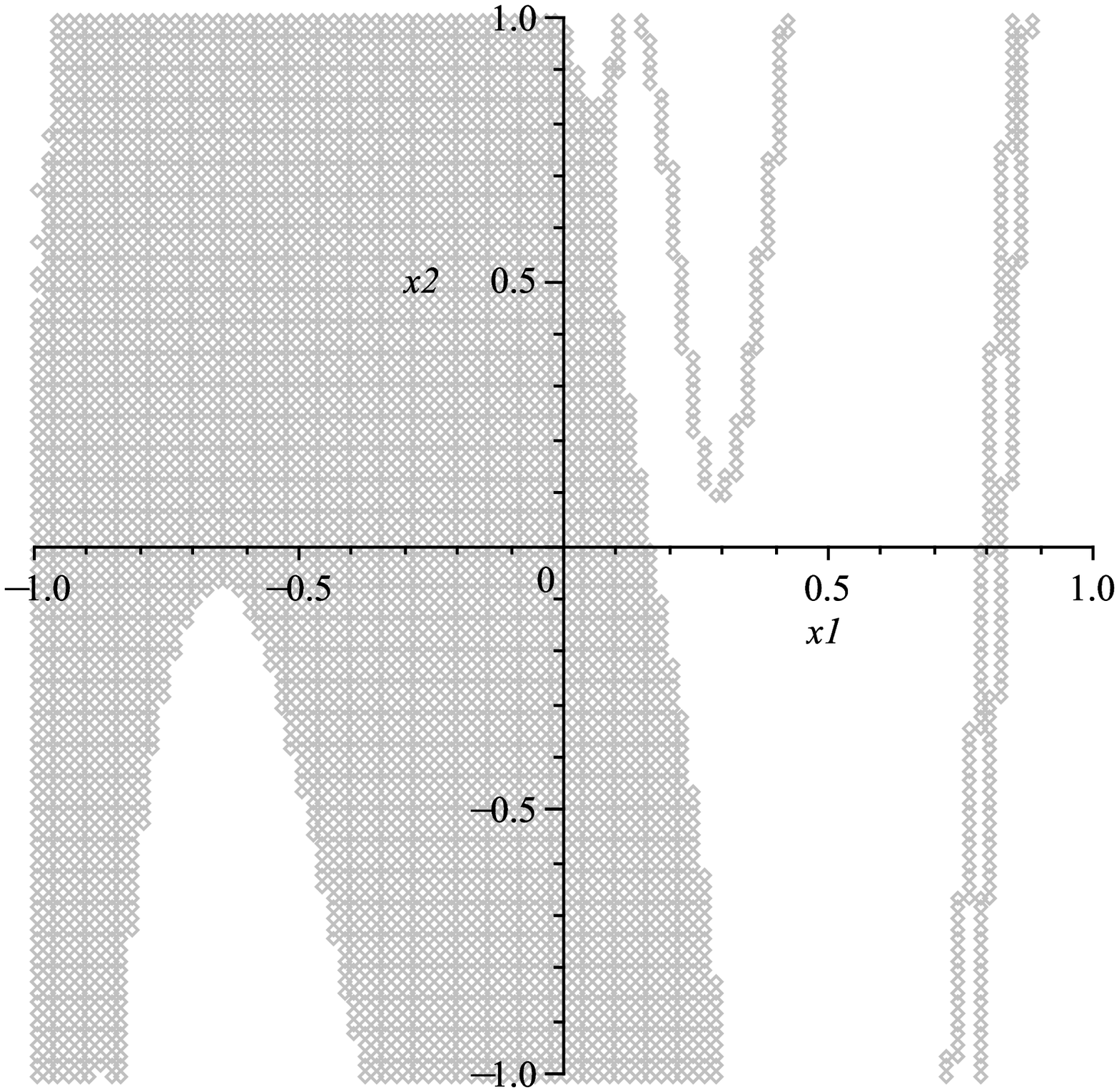}}}
 \subfigure[$\delta = 10^{15}$]{\resizebox{5cm}{5cm}{\includegraphics{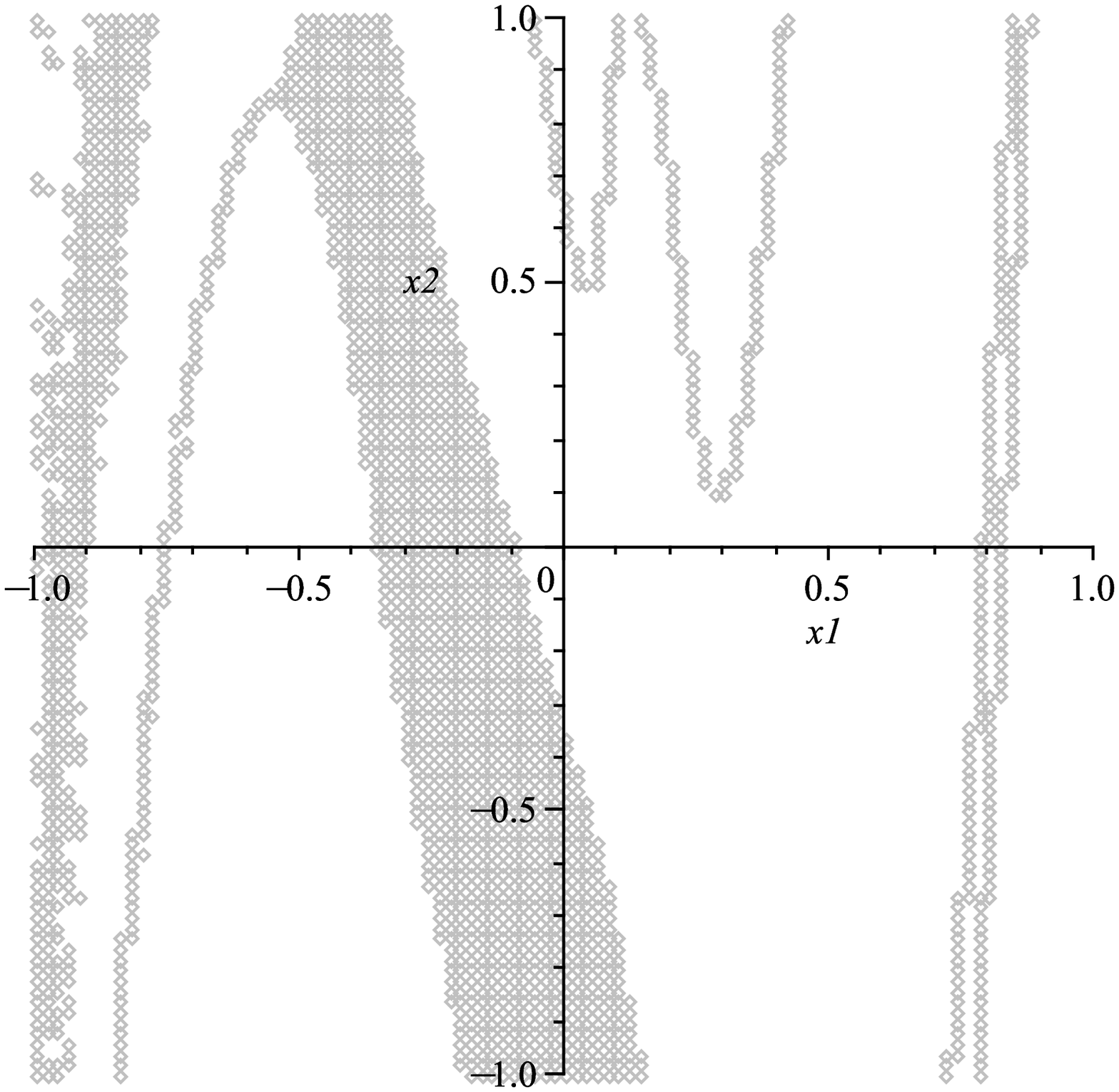}}}
 \subfigure[$\delta = 10^5$]{\resizebox{5cm}{5cm}{\includegraphics{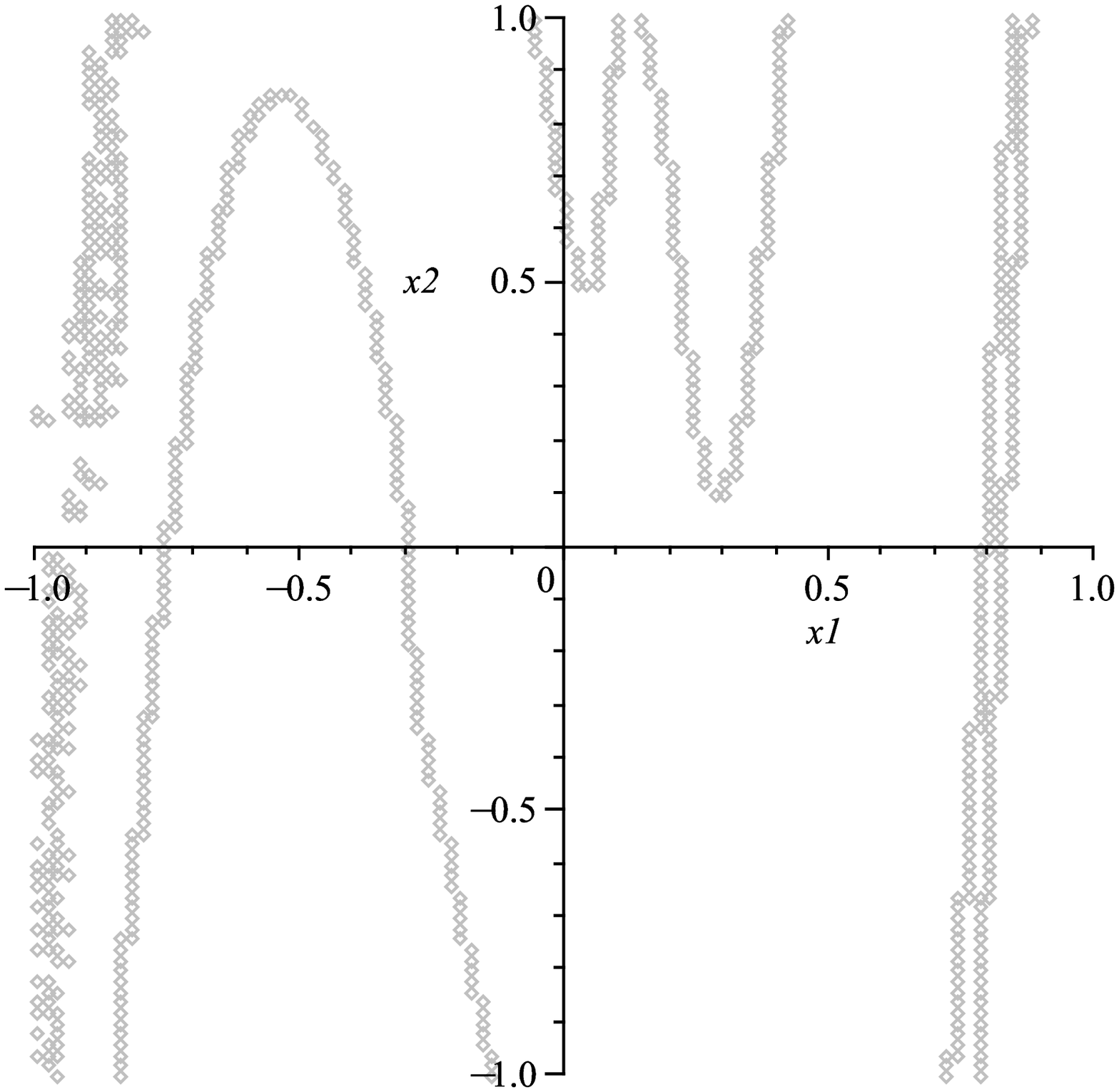}}}
\caption{Observability tests for Hindmrsh-Rose model neuron (\ref{eq:HR})}\label{fig:ObservabilityMaps}
\end{figure}

To test if there are points of local observability of system (\ref{eqn_hindmarsh_rose}) it is sufficient to find a point in the state space
of (\ref{eqn_hindmarsh_rose}) at which the rank condition
(\ref{eq:observability_condition}) holds.
 Here we computed the determinant:
 \[
 \begin{split}
  D(\tilde{x})&=\frac{\partial}{\partial \tilde{x}}
 \left(
   h(\tilde{x})\quad L_f h(\tilde{x})\quad L^2_f h(\tilde{x})\quad ...\quad L^{n-1}_fh(\tilde{x})
 \right)^T\\
 \tilde{x}&=(x_1,x_2,{\theta}_{13}, {\theta}_{12}, \dot{\theta}_{11}, {\theta}_{10}, {\theta}_{22}, \theta_{21}, {\lambda})^{T}
 \end{split}
 \]
on a sparse grid (of $101 \times 101$ pixels) and plotted those
regions for which the determinant
is less than a certain value, $\delta$. The neuron parameters
were set to $L=-2, \theta_{13}=-10, \theta_{12}=-4,
\theta_{11}=6, \theta_{10}=1, \theta_{22}=-32,
\theta_{21}=-32$. Figure \ref{fig:ObservabilityMaps} shows
results (obtained using Maple) for various values of $\delta$.
The shaded regions correspond to the domains where
$D(\tilde{x})<\delta$. According to these results, when the
value of delta is made sufficiently small, condition
$D(\tilde{x})>\delta$ holds for almost all points in the grid.
This suggests that there are domains in which model (\ref{eq:HR}) is indeed at least locally observable.

\begin{figure}
 \centering
 \subfigure[$\delta = 10000$]{\resizebox{5cm}{5cm}{\includegraphics{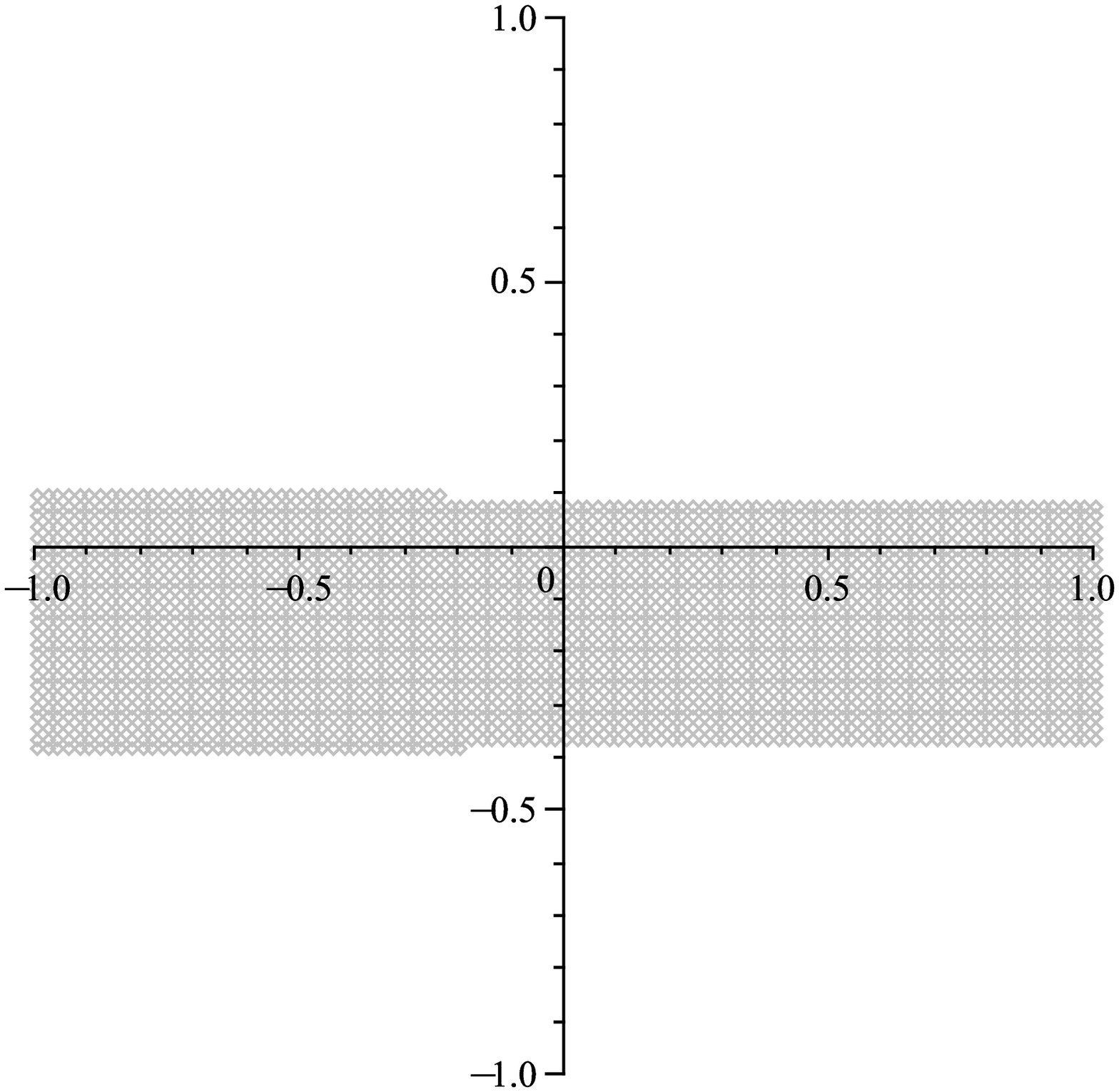}}}
 \subfigure[$\delta = 1000$]{\resizebox{5cm}{5cm}{\includegraphics{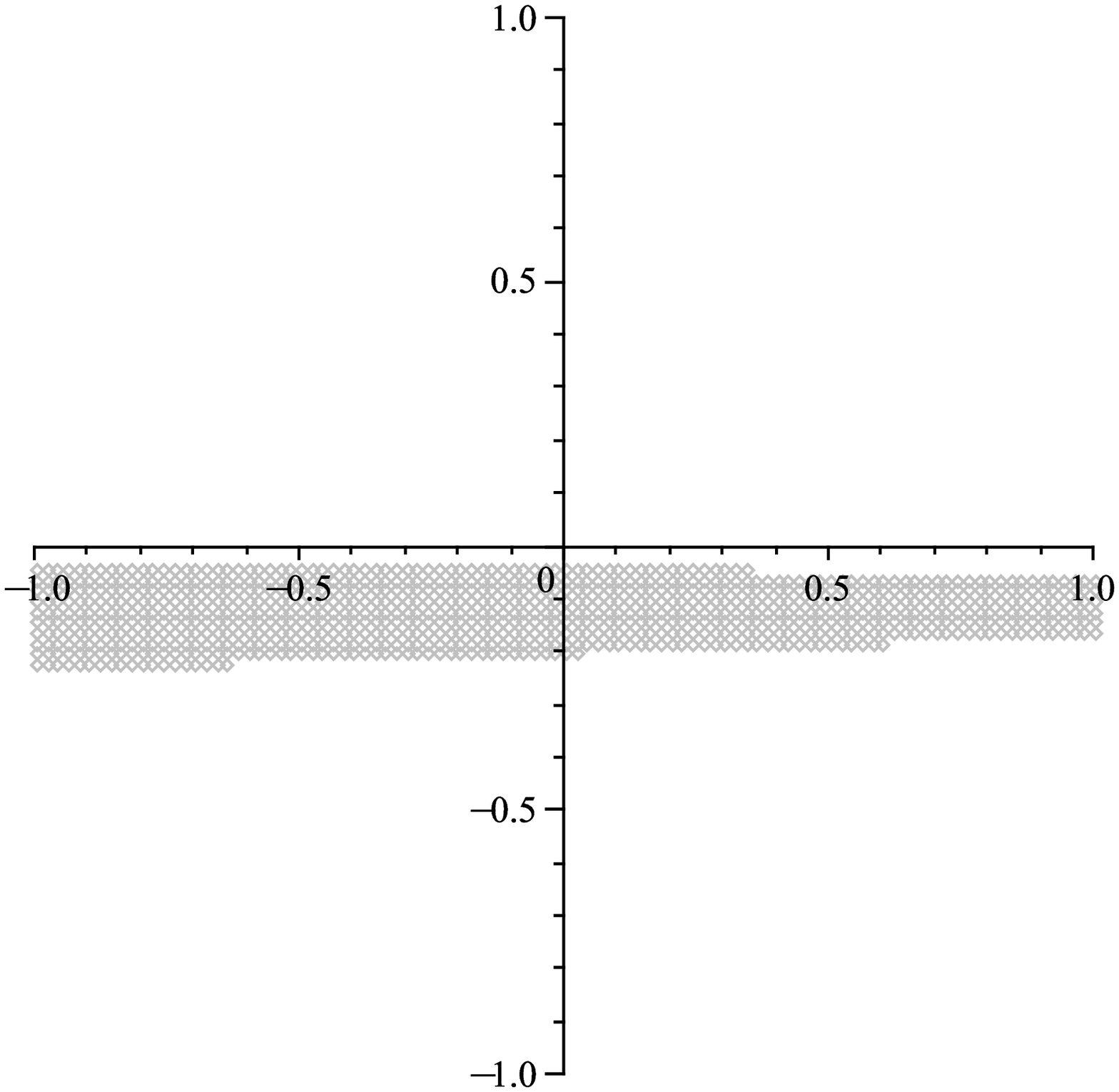}}}
 \subfigure[$\delta = 100$]{\resizebox{5cm}{5cm}{\includegraphics{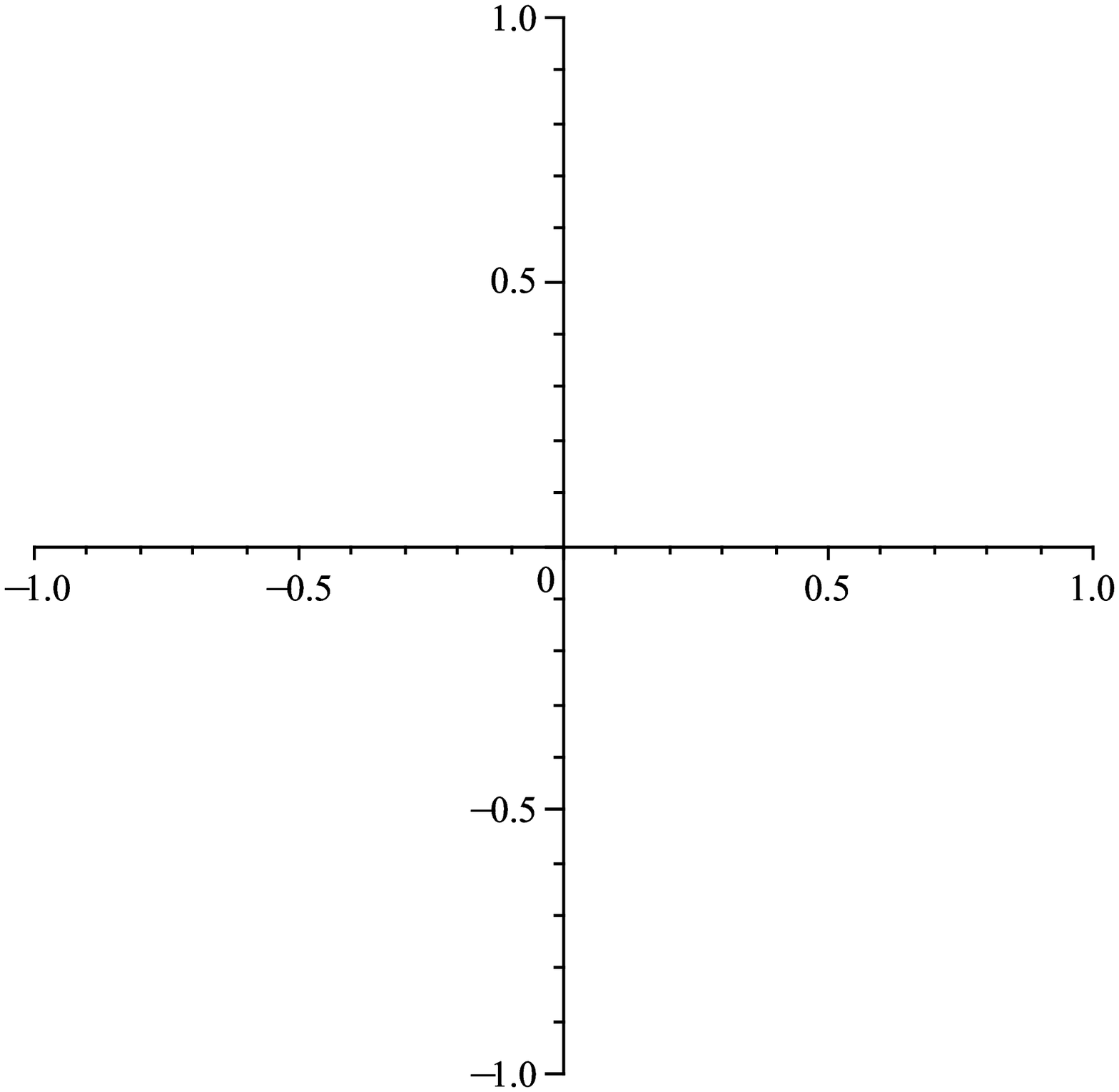}}}
\caption{Observability tests for Morris-Lecar model neuron}\label{fig:ObservabilityMapsML}
\end{figure}

Let us now consider a more realistic, with respect to
biological plausibility, set of equations. One of  the simplest
models of this type is the Morris-Lecar system
\cite{Morris_Lecar}:
\begin{eqnarray}
 \label{eqn_morris_lecar}
\begin{array}{c}
 \dot{v}(t) =
 -g_{Ca}\left(\frac{1}{2}+\frac{1}{2}\tanh\left(\frac{v(t)+1}{E_4}\right)(v(t)-E_1)\right)
 -g_K w(t)(v(t)-E_2)-g_m(v(t)-E_3)
\\
 \dot{w}(t) =
 \frac{1}{5}\left(\frac{1}{2}+\frac{1}{2}\tanh\left(\frac{v(t)+1}{E_5}\right)
 -w(t)\right)\cosh\left(\frac{v(t)}{E_6}\right)
\\
 E_1 = 100, E_2 = -70, E_3 = -50, E_4 = 15, E_5 = 30, E_6 = 60, g_{Ca} = 1.1, g_K = 2.0, g_m = 0.5
\end{array}
\end{eqnarray}
As in the previous example we extend the system state space by
considering unknown parameter as components of the extended
state vector. This extension procedure results in the following
set of equations:
\[
 \begin{split}
 &\left(
  \begin{array}{c}
   \dot{v}(t)
  \\
   \dot{w}(t)
  \\
   \dot{g}_{Ca}(t)
  \\
   \dot{g}_K(t)
  \\
   \dot{g}_m(t)
  \\
   \dot{\lambda}(t)
  \end{array}
  \right)
 =
 \left(
  \begin{array}{c}
   -g_{Ca}\left(\frac{1}{2}+\frac{1}{2}\tanh\left(\frac{v+1}{15}\right)(v-100)\right)
   -g_k w (v+70)-g_m(v+50)
  \\
   \frac{1}{5}\left(\frac{1}{2}+\frac{1}{2}\tanh\left(\frac{v+1}{30}\right)
    -w\right)\cosh\left(\frac{v}{60}\right)
  \\
   0
  \\
   0
  \\
   0
  \\
   0
  \end{array}
  \right)
 \end{split}
\]
For this extended set of equations  we estimated the regions
where value of $D(x)$ exceeds some given $\delta>0$. These
regions for different values of $\delta$  are presented in
figure \ref{fig:ObservabilityMapsML}
These results demonstrate that the Morris-Lecar system
(\ref{eqn_morris_lecar}) is also locally observable.

As we have seen above, a fairly wide class of canonical
mathematical and conductance-based models of evoked responses
in neural membranes satisfy local observability conditions.
We may thus expect to be able to solve the reconstruction problem for these models.
In fact, as we show
below in Sections 3,
4, the reconstruction problem can indeed be
resolved efficiently at least for a part of unmeasured variables of the system. However, before we proceed with detailed
description of these reconstruction algorithms, let us first
review classes of systems for which solutions to the problem of
exponentially fast reconstruction of {\it all components} of
state and parameter vectors are already available in the
literature.

\subsection{Bastin-Gevers canonical form}\label{sec:BG_observer}

We start with a class of systems comprising of a linear
time-invariant part of which the equations are known and an
additive time-varying component with linear parametrization.
Parameters of this time-varying component are assumed to be
uncertain. This class of systems was presented by G. Bastin and
M. Gevers in 1989, \cite{BastinGevers:1988},  and its general
form is as follows:
\begin{equation}
\begin{split}
 \label{BG_form}
  \dot{x} &=  R   x  +   \Omega(t) \theta  +  g(t)\\
 R &=  \left(   \begin{array}{cc}    0 & k^T   \\    0 & F   \end{array}  \right)\
\Omega(t)  =  \left(   \begin{array}{c}    \Omega_1(t)   \\    \overline{\Omega}(t)   \end{array}  \right) \\
 y(t)& = x_1(t)
\end{split}
\end{equation}
In (\ref{BG_form}), $x \in \mathbb{R}^n$ is the state vector
with $y=x_1$ assigned to be the output.
 $\theta \in \mathbb{R}^p = (\theta_1,\cdot\cdot\cdot,\theta_p)^T$ is the vector of unknown parameters.
$R$ is a known matrix of constants where $k^T =
(k_2,\cdot\cdot\cdot,k_n)$ and $F$ has dimension
 $(n-1) \times (n-1)$ with eigenvalues in the open left half plane.
 $\Omega(t) \in \mathbb{R}^{n\times p}$ is an $n \times p$ matrix
 of known functions of $t$; the first row is designated $\Omega_1$ and the remaining $n-1$ rows $\overline{\Omega}$.
The vector function $g(t):\mathbb{R}\to \mathbb{R}^n$ is known.

Equations (\ref{BG_form}) are often referred to as an adaptive
observer canonical form. This is because, subject to some mild
non-degeneracy conditions, it is always possible to reconstruct
the vector of unknown parameters $\theta$ and state $x$ from
observations of $y$ over time. Moreover, the reconstruction can
be made exponentially fast. Shown below is the adaptive
observer presented in \cite{BastinGevers:1988}. The system to
be observed, state estimator, parameter adaption, auxiliary
filter and regressor are given in equations (\ref{BG_form}),
(\ref{BG_form_estimator}),
 (\ref{BG_form_adaption}), (\ref{BG_form_aux}), (\ref{BG_form_regressor}) respectively
\begin{eqnarray}
 \label{BG_form_estimator}
  \dot{\hat{x}}(t)
 &=&
  R
   \hat{x}(t)
 +
  \Omega(t)
 \hat{\theta}
 +
 g(t)
 +
 \left(
  \begin{array}{c}
   c_1 \tilde{y}
  \\
   V(t)\dot{\hat{\theta}}
  \end{array}
 \right)
\\
\label{BG_form_adaption}
 \dot{\hat{\theta}}(t) &=& \Gamma \varphi(t) \tilde{y}(t)
\\
\label{BG_form_aux}
 \dot{V}(t) &=& F V(t) + \overline{\Omega}(t), \quad V(0) = 0
\\
\label{BG_form_regressor}
 \varphi(t) &=& V^T(t)k + \Omega_1^T(t)
\end{eqnarray}
The output is $y = x_1$, its estimate is $\hat{y} = \hat{x}_1$
and its error is $\tilde{y} = y - \hat{y}$. This observer
contains some parameters of its own which are at the design's
disposal. $\Gamma=\Gamma^{T}$ is an arbitrary positive definite
matrix, normally chosen as $\Gamma =
\textrm{diag}(\gamma_1,\gamma_2,...,\gamma_p)$, $\gamma_i > 0$.
$c_1 > 0$. The auxiliary filter $V(t)$ is an $(n-1) \times p$
matrix and $\varphi(t)$ is a $p$ vector.

Using the transformation (\ref{eqn:BG_aux_err}), the error
system (\ref{eqn:BG_err_sys}) is obtained.
\begin{eqnarray}
\label{eqn:BG_aux_err}
\begin{array}{c}
 \tilde{x}^* = \tilde{x} -
 \left(
  \begin{array}{c}
   0
\\
   V\tilde{\theta}
  \end{array}
 \right), \ \tilde{x}=x-\hat{x}
\end{array}
\end{eqnarray}

\begin{eqnarray}
\label{eqn:BG_err_sys}
\begin{array}{c}
 \dot{\tilde{x}}^*
 =
 \left(
  \begin{array}{cc}
   -c_1 & k^T
\\
   0 & F
  \end{array}
 \right)
 \tilde{x}^* +
 \left(
  \begin{array}{c}
   \varphi^T\tilde{\theta}
\\
   0
  \end{array}
  \right)
\\
 \dot{\tilde{\theta}} = -\Gamma\varphi\tilde{x}_1^*
\end{array}
\end{eqnarray}

It is shown in \cite{BastinGevers:1988}, for constant unknown
parameters, that the solution $x(t)=\hat{x}(t)$,
$\theta=\hat{\theta}(t)$ of the extended system
(\ref{BG_form_estimator}),
 (\ref{BG_form_adaption}), (\ref{BG_form_aux}), (\ref{BG_form_regressor}) is globally
exponentially stable provided certain conditions
on the regressor vector, $\varphi(t)$, are met.  These
conditions are:
\begin{itemize}
\item the regressor vector $\varphi(t)$ is bounded for all
    $t\geq 0$
\item $\dot{\varphi}(t)$ is bounded for all $t\geq 0$
    except possibly at a countable number of points
    $\{t_i\}$ such that $\mathrm{min}|t_i-t_j| \geq
    \triangle > 0$ for some arbitrary fixed $\triangle$.
\item $\varphi(t)$ is persistently exciting: that is, there
    exists positive constant $\alpha,T$ such that for all
    $t_0\geq 0$
\begin{eqnarray}\label{eq:PE_standard}
 \int_{t_0}^{t_0+T} \varphi(t) \varphi^T(t) \,\mathrm{d} t \geq \alpha I > 0
\end{eqnarray}
\end{itemize}

Formally, asymptotic properties of observer
(\ref{BG_form_estimator}), (\ref{BG_form_adaption}) are
specified in the theorem below
\cite{BastinGevers:1988}\footnote{Here we provide a slightly
reduced formulation of the main statement of
\cite{BastinGevers:1988} corresponding to the case in which the
values of $\theta$ do not change over time.}
\vspace*{0.25cm}
\begin{theorem}\label{theorem:BG} Suppose that
\begin{itemize}
\item[1) ] $c_1>0$ and $F$ is a Hurwitz matrix, that is its
    eigenvalues belong to the left half of the complex
    plane;
\item[2) ] the function $\varphi(t)$ is globally bounded in
    $t$, and its time derivative exists and is globally
    bounded for all $t\geq 0$;
\item [3) ] the function $\varphi(t)$ is persistently
    exciting.
\end{itemize}
Then the origin of (\ref{eqn:BG_err_sys}) is globally
exponentially asymptotically stable.
\end{theorem}

Adaptive observer canonical form (\ref{BG_form}) applies to
systems in which  the regressor $\Omega(t)\theta$ does not
depend explicitly on the unmeasured components of the state
vector. The question, however, is when a rather general
nonlinear system can be transformed into the proposed canonical
form. This question was addressed in \cite{MarinoTomei:1992} in
which a modified adaptive observer canonical form was proposed
together with necessary and sufficient conditions describing
when a given system can be transformed into such form via a
diffeomorphic coordinate transformation. This canonical form is
described in the next subsection.

\subsection{Marino-Tomei canonical form}

The canonical form presented in \cite{MarinoTomei:1992} is now
shown here. The system to be observed (\ref{MT_form}), state
estimator (\ref{MT_estimator}) and parameter adaption
(\ref{MT_adaption})
 are given below
\begin{eqnarray}
\begin{array}{rcl}
 \label{MT_form}
  \dot{x}(t)
 &=&
  A_1 x(t)
 +
  \phi_0(y(t),u(t)) 
 +
 b\sum_{i=1}^p\beta_i(y(t),u(t))\theta_i
\\
 y(t) &=& C_1x(t)
\\
 A_1 &=&
 \left(
  \begin{array}{ccccc}
   0 & 1 & 0 & ... & 0
  \\
   0 & 0 & 1 & ... & 0
  \\
   . & . & . & ... & 0
  \\
   0 & 0 & 0 & ... & 1
  \\
   0 & 0 & 0 & ... & 0
  \end{array}
  \right)
\\
 C_1 &=&
 \left(
  \begin{array}{ccccc}
   1 & 0 & 0 & ... & 0
  \end{array}
  \right)
\\
 && x(t) \in \mathbb{R}^n, 
 y(t) \in \mathbb{R}, \beta_i(\cdot,\cdot):\mathbb{R}\times\mathbb{R}\to\mathbb{R}
\end{array}
\end{eqnarray}
In (\ref{MT_form}) $x(t) \in \mathbb{R}^n$ is the state vector
with $x_1(t)$ assigned to be the output $y$.
 Matrices $A_1,C_1$ are in canonical observer form.
 $\theta \in \mathbb{R}^p = (\theta_1,\cdot\cdot\cdot,\theta_p)^T$ is the vector of unknown parameters.
 The functions $\beta_i$ are known, bounded and piecewise continuous functions of $y(t),u(t)$.
 The column vector $b\in \mathbb{R}^n$ is assumed to be Hurwitz\footnote{We say that a vector $b=(b_1,\dots,b_n)^{T}\in\Real^n$ is Hurwitz if all roots of the corresponding polynomial $b_1 p^{n-1}+\cdots+ b_{n-1} p+ b_n$ have negative real part.} with $b_1\ne 0$.

Shown below is the adaptive observer presented in
\cite{MarinoTomei:1992}
\begin{eqnarray}
 \label{MT_estimator}
  \dot{\hat{x}}(t)
 &=&
  (A_1-KC_1) \hat{x}(t)
 +
  \phi_0(y(t),u(t))
 +
 b\sum_{i=1}^p\beta_i(y(t),u(t))\hat{\theta}_i
 +
 Ky(t)
\\
 \label{MT_adaption}
 \dot{\hat{\theta}} &=& \Gamma \beta(t)(y-C_1\hat{x}) \mathrm{sign}(b_1)
\\
 K &=& \frac{1}{b_n}(A_1 b+\lambda b) = (k_1,\cdot\cdot\cdot,k_n)^T
\end{eqnarray}
with $\Gamma$ an arbitrary symmetric positive definite matrix and
$\lambda$ an arbitrary positive real. The $n \times 1$ vector, $b$,
is Hurwitz with $b_n \ne 0$.

For the more general case where the vector $b$ is an arbitrary vector, an observer is presented in \cite{MarinoTomei:1995}.
\vspace*{0.25cm}
\begin{theorem} \cite{Marino90}
\label{Marino_Thm} There exists a local change of coordinates,
$z=\Phi(x)$, transforming
\begin{eqnarray}
 \dot{x} &=&
 f(x)
 +
 \sum_{i=1}^p \theta_i(t) q_i(x)
 ,\quad y = x_1
\nonumber\\
 && x \in \mathbb{R}^n, y\in \mathbb{R},
\theta_i\in\mathbb{R}, q_i:\mathbb{R}^n\to\mathbb{R}^n, n\geq 2
\end{eqnarray}
with $h(x^o)=0$ and $(f,h)$ an observable pair, into the system
\begin{eqnarray}
 \dot{z} &=&
 A_1 z
 +
 \psi(y)
 +
 \sum_{i=1}^p
  \theta_i(t) \psi_i(y)
 ,\quad y = C_1 z 
\nonumber\\
 && z\in \mathbb{R}^n
 , \psi_i:\mathbb{R}\to\mathbb{R}^n
\end{eqnarray}
with $(A_1,C_1)$ in canonical observer form (\ref{MT_form}), if
and only if
\begin{itemize}
\item[(i)] $[ad_f^ig,ad_f^jg]=0,\quad 0\leq i,j\leq n-1$
\item[(ii)] $[q_i,ad_f^jg]=0,\quad 0\leq j\leq n-2,\quad
    1\leq i \leq p$
\end{itemize}
where the vector field, $g(x)$, is uniquely defined by
\begin{eqnarray}
 \label{vector_g}
 \left\langle
\frac{\pd}{\pd x} 
 \left(
 \begin{array}{c} h(x) \\ L_fh(x) \\ ... \\ L_f^{n-1}h(x)
\end{array}
 \right)
,
 g(x)
 \right\rangle
 =
 \left( \begin{array}{c} 0 \\ 0 \\ ... \\ 1 \end{array} \right)
\end{eqnarray}
\end{theorem}

The proof of this result is made along the following lines.
Suppose we use the change of coordinates:
$z=\Phi(x)$, then we have
\begin{eqnarray}
 \dot{z} &=&
 \frac{\partial \Phi}{\partial x} \dot{x}
\\
 &=&
 \left(\frac{\partial \Phi}{\partial x} f(x)\right)_{x=\Phi^{-1}(z)} +
 \frac{\partial \Phi}{\partial x} \sum_{i=1}^p \theta_i q_i(x)
\end{eqnarray}
It is shown in \cite{Marino90} that providing we meet the
constraint:
\begin{eqnarray}
 [ad_f^ig,ad_f^jg] = 0, \quad \forall x\in U, 0\leq i,j\leq n-1 
\end{eqnarray}
then we can cast the system into the adaptive observer
canonical form
\begin{eqnarray}
 \dot{z} &=&
  A_1 z
  +\psi(y)
  +\sum_{i=1}^p \frac{\partial \Phi}{\partial x} \theta_i q_i(x)
,\quad y = C_1z
\end{eqnarray}
with $A_1,C_1$ in canonical observer form (\ref{MT_form}).
Furthermore, it is shown in \cite{Marino90} that providing we
meet the constraint
\begin{eqnarray}
 [q_i,ad_f^jg] = 0, \quad \forall x\in U, 0\leq j\leq n-2
\end{eqnarray}
then we can put the system into
\begin{eqnarray}
 \dot{z} &=&
 A_1 z
 +
 \psi(y)
 +
 \sum_{i=1}^p
  \theta_i \psi_i(y)
 ,\quad y = C_1 z
\end{eqnarray}
This representation is linear in the unknown variables,
$z_1(t),z_2(t),...,z_n(t)$, and
$\theta_1(t),\theta_2(t),...,\theta_p(t)$, while it is
nonlinear only in the output, $y(t)$, which is available for
measurement.

\vspace*{0.5cm}
\setcounter{equation}{0}
\section{Feasibility of conventional adaptive observer canonical forms}\label{sec:observer_feasibility}

In this section we consider technical difficulties preventing
straightforward
 application of conventional adaptive observers for solving the state and
parameter reconstruction problems for typical neural
oscillators. We start with the most simple polynomial systems
such as the Hindmarsh-Rose equations. We show that even for
this relatively simple class of linearly parameterized models
the problem of reconstructing all parameters of the system is a
difficult theoretical challenge. Whether complete
reconstruction is possible depends substantially on what part
of the system's right-hand side is corrupted with
uncertainties. Despite in the most general case reconstruction
of all components of the
 parameter vector by using standard techniques
may not be possible, in some special yet relevant cases
estimation of {\it a part } of the model parameters is still
achievable in principle.

Let us consider, for example, the  problem of fitting
parameters of the conventional Hindmarsh-Rose oscillator to
measured data. In particular we wish to be able to model a
single spike from the measured train of spikes evoked by a
constant current injection. Classical two-dimensional
Hindmarsh-Rose model is defined by the following system:
\begin{equation}\label{eq:HR:1}
\begin{split}
\dot{x}&=-a x^3 + b x^2 +  y + I\\
\dot{y}&= c - d x^2 -y, \ a=1, \  b=3, \ c=1, \ d=5,
\end{split}
\end{equation}
in which $I\in\mathbb{R}$ stands for the stimulation current.
Trajectories $x(t)$ of this model are known to be able to
reproduce a wide range of typical responses of actual neurons
qualitatively. Quantitative modelling, however, requires the
availability of a linear transformation of $(x(t),y(t))$ so the
amplitude and the frequency of oscillations $x(t)$ can be made
consistent with data.

In what follows we will consider (\ref{eq:HR:1}) subject to the
following class of transformations:
\begin{equation}\label{eq:HR:transformation}
\left(\begin{array}{cc}
x_1\\
x_2
\end{array}\right)=\left(\begin{array}{cc} k_1 & 0 \\ k_2 & k_3 \end{array}\right) \left(\begin{array}{cc}
x\\
y
\end{array}\right) + \left(\begin{array}{cc}
p_x\\
p_y
\end{array}\right), \ k_2<k_3
\end{equation}
where $k_i>0$ and $p_x$, $p_y\in\mathbb{R}$ are unknown.
Transformations (\ref{eq:HR:transformation}) include stretching and translations as a special case. 
 In addition to (\ref{eq:HR:transformation}) we will also allow that the time constants in the right-hand
side of (\ref{eq:HR:1}) be slowly time-varying.
This will allow us to adjust scaling of the system trajectories
with respect to time.

Taking these considerations into account we obtain the
following re-parameterized description of model
(\ref{eq:HR:1}):
\begin{equation}\label{eq:HR:2}
\begin{split}
\dot{x}_1&=\sum_{i=0}^3 x_1^i \theta_{1,i} + x_2 \\
\dot{x}_2&= -\lambda x_2 + \sum_{i=1}^3 x_1^i \theta_{2,i}
\end{split}
\end{equation}
Alternatively, in vector-matrix notation we obtain:
\begin{equation}\label{eq:HR:3}
\left(
\begin{array}{c}
\dot{x}_1\\
\dot{x}_2
\end{array}\right)
= A(\lambda)
\left(
\begin{array}{c}
{x}_1\\
{x}_2
\end{array}\right) + \varphi(x_1)\theta
\end{equation}
where
\begin{equation}
\label{eq:HR:matrix}
\begin{array}{c}
A(\lambda)=\left(\begin{array}{cc} 0 & 1\\ 0 & -\lambda\end{array}\right), \
\varphi(x_1)=\left(\begin{array}{ccccccc}
  x_1^3 & x_1^2 & x_1 & 1 & 0 & 0& 0\\
  0     &  0    &  0  & 0 & x_1^3 & x_1^2 & x_1
\end{array}\right)
\\
 \theta = (\theta_{1,3},\theta_{1,2},\theta_{1,1},\theta_{1,0},\theta_{2,3},\theta_{2,2},\theta_{2,1})
\end{array}
\end{equation}

One of the main obstacles is that the original equations of
neural dynamics
 are not written in any of the canonical forms for which the reconstruction
algorithms are available. The question, therefore, is if there
exists an invertible coordinate transformation such that the
model equations can be rendered canonic. Below we demonstrate
that this is generally not the case if the transformation is
parameter-independent. This is formally stated in Section
3.1. However, if we allow our
transformation to be both parameter and time-dependent,
 a relevant class of models with polynomial right-hand sides can be transformed
into one of the canonic forms. This is demonstrated in Section
3.2.

\subsection{Parameter-independent time-invariant transformations}\label{sec:Parameter-independent}

Let us consider a class of systems that can be described by
(\ref{eq:HR:3}). Clearly this system is not in a canonical
adaptive observer form because $A(\lambda)$ depends on the
unknown parameter $\lambda$ explicitly. The question, however,
is if there exists a differentiable coordinate transformation
\[
z=\Phi(x), \ z_1=x_1
\]
such that in the new coordinates the equations of system
(\ref{eq:HR:3})
 satisfy one of the canonic descriptions. We show that the answer to this
 question is negative, and it follows from the following slightly more general statement
\vspace*{0.25mm}
\begin{theorem}\label{theorem:parameter:independent} The system
\begin{eqnarray}
 \dot{x} &=&
f(x)
 +
 \sum_{i=1}^p
  \theta_i q_i(x)
 ,\quad y = h(x)
\nonumber\\
 && x \in \mathbb{R}^n, y\in \mathbb{R},
q_i:\mathbb{R}^n\to\mathbb{R}^n, n\geq 2
\end{eqnarray}
with
\begin{eqnarray}
 f(x) =
 \left(
  \begin{array}{ccccc}
   0 & 1 & 1 & ... & 1
\\
   0 & 0 & 0 & ... & 0
\\
   . & . & . & ... & .
\\
   0 & 0 & 0 & ... & 0
 \end{array}
 \right)x,\quad h(x) =
 \left(
  \begin{array}{ccccc}
   1 & 0 & 0 & ... & 0
 \end{array} \right)x
\end{eqnarray}
cannot be transformed by diffeomorphic 
change of coordinates, $z = \phi(x)$, into
\begin{eqnarray}
 \dot{z} &=&
 A_1z
 +
 \psi_0(y)
 +
 \sum_{i=1}^p
  \theta_i \psi_i(y)
 ,\quad y = C_1z 
\nonumber\\
 && z\in \mathbb{R}^n, y\in \mathbb{R},
\psi_i:\mathbb{R}\to\mathbb{R}^n
\end{eqnarray}

with $A_1,C_1$ in canonical observer form (\ref{MT_form}), if
either (i) $n > 2$ or (ii) there exists $i \in \{1,...,p\}$,
$j\in \{2,...,n\}$ such that $\partial q_i / \partial x_j \neq
0$.
\end{theorem}
The proof of Theorem \ref{theorem:parameter:independent} and
other results are provided in the Appendix.

\subsection{Parameter-dependent and time-varying transformations}\label{sec:Parameter-dependent}

Let us now consider the case in which the transformation
$z=\Phi(x,\lambda,\theta,t)$ is allowed to depend on unknown
parameters and time. As we show below, this class of
transformations is much more flexible. In principle it allows
us to solve the problem of {\it partial} state and parameter
reconstruction for an important class of oscillators with
polynomial right-hand side and time-invariant time constants.

We start by searching for a transformation $\Phi$:
\[
\Phi: \ q=T(\lambda)x, \ |T(\lambda)|\neq 0
\]
such that
\begin{equation}\label{eq:transformation:1}
T(\lambda)A(\lambda)T^{-1}(\lambda)=\left(\begin{array}{cc} \star & 1 \\ \star & 0 \end{array}\right)
\end{equation}
where the matrix $A(\lambda)$ is defined as in
(\ref{eq:HR:matrix}). It is easy to see that the transformation
satisfying this constraint exists, and it is determined by
\begin{equation}\label{eq:transformation:2}
T(\lambda)= \left(\begin{array}{cc} 1 & 0 \\ \lambda & 1 \end{array}\right).
\end{equation}

According to (\ref{eq:transformation:1}),
(\ref{eq:transformation:2}) and (\ref{eq:HR:matrix}) equations
of (\ref{eq:HR:3}) in the coordinates $q$ can be written as
\begin{equation}\label{eq:HR:4}
\dot{q}= A_{1} q  + \psi(q_1) \eta(\theta,\lambda),
\end{equation}
where
\begin{equation}\label{eq:HR:4:matrix}
\begin{split}
A_1&=\left(\begin{array}{cc} 0 & 1 \\ 0 & 0 \end{array}\right), \ \psi(q_1)=\left(\begin{array}{cccccccc}
  q_1^3 & q_1^2 & q_1 & 1 & 0 & 0& 0 & 0\\
  0     &  0    &  0  & 0 & q_1^3 & q_1^2 & q_1 & 1
\end{array}\right),\\
\eta(\theta,\lambda)&=(\theta_{1,3},\theta_{1,2},\theta_{1,1}-\lambda,\theta_{1,0},\lambda \theta_{1,3}+\theta_{2,3},\lambda \theta_{1,2}+\theta_{2,2}, \lambda \theta_{1,1}+\theta_{2,1},\lambda\theta_{1,0})^{T}
\end{split}
\end{equation}
\vspace*{0.25cm}
\begin{remark}\label{rem:reconstruct:1} Notice that
\begin{itemize}
\item availability of the parameter vector $\eta$ in
    (\ref{eq:HR:4}), (\ref{eq:HR:4:matrix}), expressed as a
    function of $\theta$, $\lambda$, implies the
    availability of $\theta$, $\lambda$ if
    $\theta_{1,0}\neq 0$. Indeed, in this case the value of
    $\lambda=\eta_{8}/\eta_4$ and the values of all
    $\theta_{i,j}$ are uniquely defined by $\eta_i$;
\item condition $\theta_{1,0}\neq 0$ is sufficient for
    reconstructing the values of $x_2$ provided that $q$
    and $\eta$ are available; indeed in this case
    $x=T^{-1}(\lambda)q$
\end{itemize}
\end{remark}

As follows from Remark \ref{rem:reconstruct:1} the problem of
state and  parameter reconstruction of (\ref{eq:HR:3}) from
measured data $x_1(t)$ amounts to solving the problem of state
and parameter reconstruction of (\ref{eq:HR:4}). In order to
solve this problem we shall employ yet another coordinate
transform:
\begin{equation}\label{eq:transformation:3}
\begin{split}
z_1&=q_1\\
z_2&= q_2 + \zeta^{T}(t) \eta
\end{split}
\end{equation}
in which the functions $\zeta(t)$ are some differentiable
functions of time.
 Coordinate transformation (\ref{eq:transformation:3}) is clearly time-dependent.
 The role of this additional transformation is to transform the equations
of system (\ref{eq:HR:4}) into the form for which a solution
already exists.

Definitions  of these functions, specific estimation algorithms
and their convergence properties are discussed in detail in the
next section.

\subsection{Observers for transformed equations}

\subsubsection{Bastin-Gevers Adaptive Observer}
Proceeding from (\ref{eq:HR:4}), (\ref{eq:HR:4:matrix}) and
applying a second change of coordinates given by
\begin{eqnarray}
 \left(
  \begin{array}{c}
   z_1
\\
   z_2
  \end{array}
  \right)
 &=&
 \left(
  \begin{array}{cc}
   1 & 0
  \\
   \frac{f}{k} & \frac{1}{k}
  \end{array}
 \right)
 \left(
  \begin{array}{c}
   q_1
\\
   q_2
  \end{array}
  \right),\nonumber
\end{eqnarray}
where $f\in\Real_{<0}$ and $k\in\Real$ are some design
parameters, we obtain the canonical form (\ref{BG_form})
presented in \cite{BastinGevers:1988}
\begin{eqnarray}
\label{eqn_BG1}
\begin{array}{c}
 \left(
  \begin{array}{c}
   \dot{z}_1
\\
   \dot{z}_2
  \end{array}
  \right)
 =
 R 
 \left(
  \begin{array}{c}
   z_1
\\
   z_2
  \end{array}
 \right)
 +
 g(t)
 +
 \Omega(y) 
 \eta(\theta,\lambda)
\\
 R 
 =
 \left(
  \begin{array}{cc}
   0 & k
  \\
   0 & f
  \end{array}
 \right)
\\
 g(t) =
 -y
 \left(
  \begin{array}{c}
   f
  \\
   \frac{f^2}{k}
  \end{array}
 \right)
\\
 \Omega(y) =
 \left(
  \begin{array}{cccccccc}
   y^3 & y^2 & y & 1 & 0 & 0 & 0 & 0
\\
   \frac{f}{k}y^3 &
   \frac{f}{k}y^2 &
   \frac{f}{k}y &
   \frac{f}{k}&
   \frac{1}{k}y^3 &
   \frac{1}{k}y^2 &
   \frac{1}{k}y &
   \frac{1}{k}
 \end{array}
\right)
\\
 \eta(\theta,\lambda) =
 \left(
  \theta_{13},\,
  \theta_{12},\,
  \theta_{11}-\lambda,\,
  \theta_{10},\,
  \theta_{23}+\lambda\theta_{13},\,
  \theta_{22}+\lambda\theta_{12},\,
  \theta_{21}+\lambda\theta_{11},\,
  \lambda\theta_{10}
 \right)^T
\end{array}
\end{eqnarray}
System (\ref{eqn_BG1}) now is in the Bastin-Gevers adaptive
observer canonical form. Notice that the parameter vector
$\eta(\lambda,\theta)$ remains unchanged and recall Remark
\ref{rem:reconstruct:1}. Let us proceed to the observer
construction following the steps described in
(\ref{BG_form_estimator}) -- (\ref{eqn:BG_err_sys}).

We start by introducing an auxiliary filter of which the
general form is given by (\ref{BG_form_aux}).
 According to (\ref{eqn_BG1}) the auxiliary filter is defined as follows:
\begin{eqnarray}\label{eq:BG_HR_filter}
\begin{array}{lll}
 \dot{v}_1 &=& f v_1 + \frac{f}{k}y^3
\\
 \dot{v}_2 &=& f v_2 + \frac{f}{k}y^2
\\
 \dot{v}_3 &=& f v_3 + \frac{f}{k}y
\\
 \dot{v}_4 &=& f v_4 + \frac{f}{k}
\\
 \dot{v}_5 &=& f v_5 + \frac{1}{k}y^3
\\
 \dot{v}_6 &=& f v_6 + \frac{1}{k}y^2
\\
 \dot{v}_7 &=& f v_7 + \frac{1}{k}y
\\
 \dot{v}_8 &=& f v_8 + \frac{1}{k}
\end{array}
\end{eqnarray}
Hence in accordance with (\ref{BG_form_regressor}) the
regressor vector $\varphi(t)$ is written as
\begin{eqnarray}\label{eq:BG_HR_regressor}
\begin{array}{lll}
 \varphi_1 &=& k v_1 + y^3
\\
 \varphi_2 &=& k v_2 + y^2
\\
 \varphi_3 &=& k v_3 + y
\\
 \varphi_4 &=& k v_4 + 1
\\
 \varphi_5 &=& k v_1
\\
 \varphi_6 &=& k v_2
\\
 \varphi_7 &=& k v_3
\\
 \varphi_8 &=& k v_4
\end{array}
\end{eqnarray}
and the observer equations are as follows:
\begin{eqnarray}\label{eqn:BG_observer_sys1}
\begin{split}
 \dot{\hat{x}} & =
 \left(
  \begin{array}{cc}
   -c_1 & k
\\
   0 & f
  \end{array}
 \right)
 \hat{x} + \left(
  \begin{array}{c}
   c_1
\\
   0
  \end{array}
  \right)\hat{x}_1 + g(t) +  \Omega(x_1)\hat{\eta} + 
  \left(\begin{array}{c}
   0
\\
   V \dot{\hat{\eta}}
  \end{array}
  \right)
\\
 \dot{\hat{\eta}} & =  \Gamma\varphi(x_1-\hat{x}_1), \\
 V &=(v_1,v_2,v_3,v_4,v_5,v_6,v_7,v_8)
\end{split}
\end{eqnarray}

Taking (\ref{eq:BG_HR_filter}) -- (\ref{eqn:BG_observer_sys1}),
and (\ref{eqn:BG_err_sys}) into
 account we obtain the following equations governing the dynamics of the estimation error,
$(\tilde{x}^*,\tilde{\eta})^{T}$
\begin{eqnarray}\label{eqn:BG_err_sys1}
\begin{array}{c}
 \dot{\tilde{x}}^*
 =
 \left(
  \begin{array}{cc}
   -c_1 & k
\\
   0 & f
  \end{array}
 \right)
 \tilde{x}^* +
 \left(
  \begin{array}{c}
   \varphi^T\tilde{\eta}
\\
   0
  \end{array}
  \right)
\\
 \dot{\tilde{\eta}} = -\Gamma\varphi\tilde{x}_1^*
\end{array}
\end{eqnarray}
The auxiliary filter (\ref{eq:BG_HR_filter}) acts here as an
inherent component of a
 time-varying coordinate transformation rendering the error dynamics into  (\ref{eqn:BG_err_sys}).
 This coordinate transformation is similar to that defined by (\ref{eq:transformation:3}),
provided that $z$, $\eta$ in (\ref{eq:transformation:3}) are
replaced by estimation errors $\tilde{x}^\ast$, $\tilde{\eta}$.

Let us now explore asymptotic properties of the observer. First
we notice that $v_4(t),\, v_8(t)$ both converge to constant
values exponentially fast as $t\to\infty$.  In fact,
\[
 \lim_{t\rightarrow\infty}v_4(t)=-\frac{1}{k}, \ \lim_{t\rightarrow\infty} v_8(t) = -\frac{1}{fk}.
\]
Thus accordingly  $\varphi_4(t),\, \varphi_8(t)$ both tend to
constant values as $t\to\infty$:
\[
\lim_{t\rightarrow\infty} \varphi_4(t) =0, \ \lim_{t\rightarrow\infty} \varphi_8(t) = -\frac{1}{f}.
\]
The latter fact implies that the persistency of excitation
requirement is necessarily violated for regressor
(\ref{eq:BG_HR_regressor}). Indeed, condition
(\ref{eq:PE_standard}) does not hold if one of the components
of $\varphi(t)$ is exponentially converging to zero. The
question therefore, is if this approach can be used at all to
construct asymptotically  converging estimators of state and
parameters of (\ref{eqn_BG1}). The answer to this question is
provided in the corollary below
\vspace*{0.25cm}
\begin{corollary}\label{cor:BG_HR_observer} Consider the function
\[
\bar{\varphi}(t)=(\varphi_1(t),\varphi_2(t),\varphi_3(t),\varphi_5(t),\varphi_6(t),\varphi_7(t),\varphi_8(t))^{T}
\]
If it is 
globally bounded and persistently exciting, (\ref{eq:PE_standard}), then
the following holds along the
 solutions of (\ref{eq:BG_HR_filter}) -- (\ref{eqn:BG_observer_sys1}):
\[
\lim_{t\rightarrow\infty}\hat{x}(t)-x(t)=0, \  \lim_{t\rightarrow\infty} \hat{\eta}_i(t)-\eta_i, \ i\neq 4,
\]
and the convergence is exponential.
\end{corollary}
\vspace*{0.25cm}
\begin{remark}\label{rem:cor1} Corollary \ref{cor:BG_HR_observer} demonstrates that despite the
original result of \cite{BastinGevers:1988}, i.e. Theorem
\ref{theorem:BG}, does not apply
 to system (\ref{eqn_BG1}) directly one can still construct a reduced order observer for
this system. This reduced observer guarantees partial
reconstruction of unmeasured parameters,
 and this reconstruction is exponentially fast. To recover the true values of unknown parameters
one needs to solve the following system
\[
\begin{split}
  \eta_1&=\theta_{13} \\
  \eta_2&=\theta_{12} \\
  \eta_3&=\theta_{11}-\lambda\\
  \eta_5&= \theta_{23}+\lambda\theta_{13}\\
  \eta_6&= \theta_{22}+\lambda\theta_{12}\\
  \eta_7&= \theta_{21}+\lambda\theta_{11}\\
  \eta_8&=\lambda\theta_{10}
\end{split}
\]
for $\theta_i$, $\lambda$ taking the values of $\hat{\eta}_i$
as the estimates of $\eta_i$.
 Solution to this system may not be unique, hence the reconstruction is generally possible
only up to a certain scaling factor.
\end{remark}

Simulation results for this observer are presented in Section
5.

\subsubsection{Marino-Tomei Observer}

Let us define the vector-function $\zeta(t)$ in
(\ref{eq:transformation:3}) as follows:
\[
\dot{\zeta}_i=-k \zeta_i  + k \psi_{1,i}(q_1) - \psi_{2,i}(q_1), \ k\in\mathbb{R}_{>0}
\]
In this case we have
\[
\begin{split}
\dot{z}_2&=\sum_{i=1}^8\psi_{2,i}(q_1) \eta_i +k \left( -\sum_{i=1}^8\zeta_i\eta_i+\psi_{1,i}(q_1)\eta_i\right) - \sum_{i=1}^8\psi_{2,i}(q_1) \eta_i\\
&= k  \sum_{i=1}^8(-\zeta_i +\psi_{1,i}(q_1))\eta_i
\end{split}
\]
Hence, taking equality (\ref{eq:transformation:3}) into account
and expressing $q_2$ as
$q_2=z_2-\zeta^{T}(t)\eta(\theta,\lambda)$ we obtain
\begin{equation}\label{eq:HR:5}
\begin{split}
\dot{z}_1&= z_2 + \sum_{i=1}^8(-\zeta_i +\psi_{1,i}(q_1))\eta_i\\
\dot{z}_2&=  k  \sum_{i=1}^8(-\zeta_i +\psi_{1,i}(q_1))\eta_i
\end{split}
\end{equation}
Notice that $\psi_{1,4}(q_1)=\psi_{2,8}(q_1)=1$, hence
$-\zeta_4 +\psi_{1,4}(q_1)$ and $-\zeta_8 +\psi_{1,8}(q_1)$
converge to some constants in $\mathbb{R}$ exponentially fast
as $t\rightarrow\infty$. Moreover, the sum $-\zeta_4
+\psi_{1,4}(q_1)$ is converging to zero,
 and the sum $-\zeta_8 +\psi_{1,8}(q_1)$ is converging to $-1/k$ as $t\rightarrow\infty$.
Taking these facts into account we can conclude that system
(\ref{eq:HR:5}) can be rewritten in the following (reduced)
form
\begin{equation}\label{eq:HR:6}
\begin{split}
\dot{z}&= A_{1} z  + b \phi^T(z_1,t) {\upsilon}(\theta,\lambda) + b\varepsilon(t),\\
A_1&=\left(\begin{array}{cc} 0 & 1 \\ 0 & 0 \end{array}\right), \ b=\left(\begin{array}{c} 1 \\ k \end{array}\right)\\
\phi(z_1,t)&=\left(\begin{array}{c}
                    -\zeta_1 + \psi_{1,1}(z_1)\\
                    -\zeta_2 + \psi_{1,2}(z_1)\\
                    -\zeta_3 + \psi_{1,3}(z_1)\\
                    -\zeta_8 + \psi_{1,8}(z_1)\\
                    -\zeta_5 + \psi_{1,5}(z_1)\\
                    -\zeta_6 + \psi_{1,6}(z_1)\\
                    -\zeta_7 + \psi_{1,7}(z_1)\\
                    \end{array}\right)\\
{\upsilon}(\theta,\lambda)&=(\theta_{1,3},\theta_{1,2},\theta_{1,1},\bar{\theta}_{1,0},\lambda \theta_{1,3}+\theta_{2,3},\lambda \theta_{1,2}+\theta_{2,2}, \lambda \theta_{1,1}+\theta_{2,1})^{T},
\end{split}
\end{equation}
where $\varepsilon(t)$ is an exponentially decaying term.

System (\ref{eq:HR:6}) is clearly in the adaptive canonic
observer form. Hence it admits the following adaptive observer
\begin{equation}\label{eq:passivity_observer}
\begin{split}
\dot{\hat{z}}&= A_{1} \hat{z} + L (\hat{z}-z)  + b \phi^T(z_1,t) \hat{\upsilon}\\
L&=\left(\begin{array}{cc} - l_1 & 0 \\ - l_2 & 0 \end{array}\right), \ l_1=k+1,l_2=k\\
\dot{\hat{\upsilon}}_i&=-\gamma (\hat{z}_1-z_1) \phi_i(z_1,t), \ \gamma\in\mathbb{R}_{>0}
\end{split}
\end{equation}
of which the asymptotic properties are specified in the
following Theorem
\vspace*{0.25cm}
\begin{theorem}\label{theorem:observer:passivity}
 Let us suppose that system
(\ref{eq:HR:6}) be given and its solutions are defined for all
$t$. Then, for all initial conditions, solutions of the
combined system (\ref{eq:HR:6}), (\ref{eq:passivity_observer})
exist for all $t$ and
\[
 \lim_{t\rightarrow\infty} \hat{z}(t)-z(t)=0   
\]

Furthermore, if the function $\phi(z_1,t)$ is persistently
exciting and $z(t)$ is bounded then
\[
\lim_{t\rightarrow\infty}\hat{\upsilon}(t)-\upsilon(\theta,\lambda)=0,
\]
and the dynamics of $\hat{z}-z, \hat{\upsilon}-\upsilon$ are
exponentially stable in the sense of Lyapunov.
\end{theorem}
The proof of Theorem \ref{theorem:observer:passivity} is
provided in the Appendix.

\vspace*{0.25cm}
\begin{remark} Similar to Corollary \ref{cor:BG_HR_observer} for Bastin-Gevers observer,  Theorem \ref{theorem:observer:passivity} provides
 us with a computational scheme that, subject to that $\phi(z_1,t)$ is persistently exciting,
 can be used to estimate the values of the modified vector of uncertain
 parameters $\upsilon(\theta,\lambda)$. The question, however,
 is that if the values of $\theta$, $\lambda$ can always be restored
from $\upsilon(\theta,\lambda)$. In general, the answer to this
question is negative.
 Indeed, according to (\ref{eq:HR:4}) we have
\begin{equation}\label{eq:HR:reconstruct:eq:2}
\begin{split}
\theta_{1,3}&=\upsilon_{1}\\
\theta_{1,2}&=\upsilon_{2}\\
\theta_{1,1}-\lambda&=\upsilon_{3}\\
\bar{\theta}_{1,0}=\lambda \theta_{1,0}&=\upsilon_{4}\\
\lambda \theta_{1,3}+\theta_{2,3}&=\upsilon_5\\
\lambda \theta_{1,2}+\theta_{2,2}&=\upsilon_{6}\\
\lambda \theta_{1,1}+\theta_{2,1}&=\upsilon_{7}
\end{split}
\end{equation}
As follows from (\ref{eq:HR:reconstruct:eq:2}) one can easily
recover the values of $\theta_{1,3}$, $\theta_{1,2}$, and
$\theta_{1,1}$. However, recovering the values of  remaining
parameters explicitly from the estimates of
$\upsilon(\theta,\lambda)$ is possible only up to a certain
scaling parameter. Indeed, if the number of unknowns in
(\ref{eq:HR:reconstruct:eq:2}) exceeds the number of equations
by one.
\end{remark}
\vspace*{0.25cm}
\begin{remark}\label{rem:uniqueness:issue} Notice that in the relevant special cases,
when the value of either $\theta_{2,3}$, $\theta_{2,2}$, or
$\theta_{2,1}$ is zero,
 such reconstruction is obviously possible. Let us suppose that $\theta_{2,3}=0$.
Hence the value of $\lambda$ can be expressed from
(\ref{eq:HR:reconstruct:eq:2}) as
\begin{equation}\label{eq:HR:reconstruct:eq:3}
\lambda=\frac{\upsilon_5}{\upsilon_1},
\end{equation}
and thus the rest of parameters can  be reconstructed as well.
 Due to the presence of division in (\ref{eq:HR:reconstruct:eq:3}),
 this scheme may be sensitive  to persistent perturbations when $\upsilon_5=\lambda\theta_{1,3}$ is small.
\end{remark}

So far we considered special cases of (\ref{eq:non_canonic}) in
which the time constants of unmeasured variables were unknown
yet constant and parametrization of the right-hand side was
linear.
 As we mentioned in Remark \ref{rem:uniqueness:issue},
 even for this simpler class of systems solving the problem of
 parameter reconstruction may not be a straightforward operation. For example,
 if there are cubic, quadratic and linear terms in the second equation
of (\ref{eq:HR:3}) then recovering all parameters of
(\ref{eq:HR:3}) by
 observer (\ref{eq:passivity_observer}) may not be possible.
Nonlinear parametrization, time-varying time constants and
nonlinear coupling between equations in the right-hand side of
(\ref{eq:non_canonic})
 make the reconstruction problem even more complicated. Even though there are results that partially address  the issue of nonlinear parametrization, see e.g. \cite{tpt2003_tac}, \cite{IEEE_TAC_2007}, \cite{Cao_2003}, \cite{ALCOSP_2004}, \cite{Lin_2002_smooth}, the estimation problem for systems with general nonlinear parametrization is still an open issue.

In the next section we show that for a large subclass of
(\ref{eq:non_canonic}) there always exists an observer that
solves the problem of state and parameter reconstruction from
the measurements of $V$. Moreover the structure of this
observer does not depend significantly on specific equations
describing dynamics of the observed system. For this reason,
and similarly to \cite{Dyn_Con:Ilchman:97}, we refer to this
class of observers as {\it universal adaptive observers}.

\vspace*{0.5cm}
\setcounter{equation}{0}
\section{Universal adaptive  observers for conductance-based models}\label{sec:universal}

The ideas of universal adaptive observers for systems with
nonlinearly parameterized uncertainty was introduced in a
series of works \cite{SIAM_non_uniform_attractivity},
 \cite{Adaptive_Observers}  devoted to the study of convergence to unstable invariant sets.
 Here we provide a review of these results and discuss how they can be applied to the
problem of state and parameter reconstruction of
(\ref{eq:non_canonic}).

The following class of models is considered in
\cite{Adaptive_Observers}:
\begin{equation}\label{eq:neural_model}
\left\{\begin{array}{ll} \dot{x}_0&= \theta_0^T \ \phivec_0(x_0,p_0,t) + \sum_{i=1}^n c_i(x_0,q_i,t) x_i + c_0(x_0,q_0,t)+\xi_0(t)+ u(t)\\
\dot{x}_1&= - \beta_1(x_0,\tau_{1},t) \ x_1 + \theta_1^T \phivec_1(x_0,p_{1},t)+\xi_1(t),\\
&\vdots\\
\dot{x}_i&= - \beta_i(x_0,\tau_{i},t) \ x_i + \theta_i^T \phivec_i(x_0,p_{i},t)+\xi_i(t),\\
&\vdots\\
\dot{x}_n&= - \beta_n(x_0,\tau_{n},t) \ x_n + \theta_n^T \phivec_i(x_0,p_{n},t)+\xi_n(t),
\end{array}\right.
\end{equation}

\[
\begin{split}
y&=(1,0,\dots,0) \bfx = x_0, \ x_i(t_0)=x_{i,0}\in\Real,\\
\bfx&=\mathrm{col}(x_0,x_1,\dots,x_n),\ \theta_i=\mathrm{col}(\theta_{i,1},\dots,\theta_{i,d_i}),
\end{split}
\]
where
\[
\begin{split}
\phivec_i:& \Real\times \Real^{m_i} \times \Real_{\geq 0} \rightarrow \Real^{d_i}, \ \phivec_i\in\mathcal{C}^{0}, \
d_i, m_i\in\Natural, \ i=\{0,\dots, n\}\\
\beta_i:&\Real\times\Real\times\Real_{\geq 0}\rightarrow\Real_{>0}, \ \beta_i\in\mathcal{C}^0,  \ i=\{1,\dots, n\}\\
c_i:&\Real\times\Real^{r_i}\times\Real_{\geq 0}\rightarrow\Real, \ c_i\in\mathcal{C}^0,  \ r_i\in\Natural, \ i=\{0,\dots, n\}
\end{split}
\]
are continuous and known functions, $u:\Real_{\geq
0}\rightarrow\Real$, $u\in\mathcal{C}^{0}$ is a known function
of time modelling the control input, and $\xi_i:\Real_{\geq
0}\rightarrow\Real$, $\xi_i\in\mathcal{C}^0$ are functions that
are unknown, yet bounded. The functions $\xi_i(t)$ represent
{\it unmodeled dynamics}, external perturbations, residuals due to the coarse-graining procedures at the stage of reduction \cite{Model_reduction_AG_2006}, etc.

Variable $y$ in system (\ref{eq:neural_model}) is the {\it
output}, and the variables $x_i$, $i\geq 1$ are the components
of state $\bfx$, that are not available for direct observation.
Vectors $\theta_i\in \Real^{d_i}$ consist of  {\it linear}
parameters of uncertainties in the right-hand side of the
$i$-th equation in (\ref{eq:neural_model}). Parameters
$\tau_{i}\in\Real$, $i=\{1,\dots,n\}$ are the unknown
parameters of time-varying relaxation rates,
$\beta_i(x_0,\tau_i,t)$, of the state variables $x_i$, and
vectors $p_{i}\in\Real^{m_i}$, $q_{i}\in\Real^{r_i}$, consist
of the {\it nonlinear} parameters of the uncertainties. The
functions $c_i(x_0,q_i,t)$ are supposed to be bounded.

Notice that system (\ref{eq:neural_model}) is almost as general
as (\ref{eq:non_canonic}).
 The only difference is that variables $x_i$, $i\geq 1$ enter the first equation of (\ref{eq:neural_model}) as
\[
\sum_{i=1}^n c_i(x_0,q_i,t) x_i
\]
whereas the corresponding variables $r_i$ in system
(\ref{eq:non_canonic})
 enter the first equation in a slightly more general way
\[
 {\sum}_j \ \varphi_j(v,t) p_j(r) \theta_{j}.
\]
This difference, however, is not critical for the observers
presented in
 \cite{Adaptive_Observers} can be adjusted to deal with this more general case as well.

For notational convenience we denote:
\[
\begin{split}
\thetavec&=\col(\theta_0,\theta_1,\cdots, \theta_n), \\
\lambdavec&=\col(p_0,q_0,\tau_1, p_1,q_1 \dots,\tau_n, p_n,q_n),\\
s&=\dim{(\lambdavec)}=n+\sum_{i=0}^n (m_i + r_i) .
\end{split}
\]
Symbols $\Omega_\theta$ and $\Omega_\lambda$, respectively,
denote domains of admissible values for $\thetavec$ and
$\lambdavec$.

The system state $\bfx=\col(x_0,x_1,\cdots,x_n)$ is not
measured;  only the values of the input $u(t)$ and the output
$y(t)=x_0(t)$, $t\geq t_0$ in (\ref{eq:neural_model}) are
accessible over any time interval $[t_0,t]$ that belongs to the
history of the system. The actual values of parameters
$\thetavec$, $\lambdavec$ are assumed to be unknown {\it
a-priori}. We assume however, that they belong to a set, e.g. a
hypercube, with known bounds:
$\theta_{i,j}\in[\theta_{i,\min},\theta_{i,\max}]$,
$\lambda_i\in[\lambda_{i,\min},\lambda_{i,\max}]$.

Instead of imposing the traditional requirement of asymptotic
estimation of
 the unknown parameters with arbitrarily small error we relax our demands to
 estimating the values of state and parameters of (\ref{eq:neural_model}) up
to a certain tolerance. This is because we allow unmodeled
dynamics, $\xi_i(t)$, in the right-hand side of
(\ref{eq:neural_model}). As a result of such a practically
important addition there may exist a set of systems of which
the solutions are relatively close to the measured data
 yet
their parameters could be different. Instead of just one value
of unknown parameter vectors $\thetavec$, $\lambdavec$ we
therefore have to deal with a set of $\thetavec$, $\lambdavec$
corresponding to the solutions of (\ref{eq:neural_model}) that
over time are sufficiently close. This set of model parameters
is referred to as an {\it equivalence class} of
(\ref{eq:neural_model}).


%

Similarly to canonical observer schemes \cite{Marino90},
\cite{BastinGevers:1988}, \cite{MarinoTomei:1995} the method
presented in \cite{Adaptive_Observers} relies on the
ability to evaluate the integrals
\begin{equation}\label{eq:filtered}
\mu_i(t,\tau_i,p_i)\triangleq\int_{t_0}^t
e^{-\int_{\tau}^{t}\beta_i(x_0(\chi),\tau_{i},\chi)d\chi}\phivec_i(x_0(\tau),p_{i},\tau)d\tau
\end{equation}
at a given time $t$ and for the given values of $\tau_i$, $p_i$
within a given accuracy. In classical adaptive observer
schemes,  the values of $\beta_i(x_0,\tau_i,t)$ are constant.
This allows us to transform the original equations by a
(possibly parameter-dependent) non-singular linear coordinate
transformation, $\Phi: \ \bfx\mapsto\bfz$, $x_1=z_1$, into an
equivalent form in which the values of all time constants are
known. In the new coordinates the variables $z_2,\dots,z_n$ can
be estimated by integrals (\ref{eq:filtered}) in which the
values of $\beta_i(x_0,\tau_i,t)$ are constant and known. This
is usually done by using auxiliary linear filters. In our case,
the values of $\beta_i(x_0,\tau_i,t)$ are not constant and are
unknown due to the presence of $\tau_i$. Yet if the values of
$\tau_i$ would
 be known we could still
estimate the values of integrals (\ref{eq:filtered}) as follows
\begin{eqnarray}\label{eq:approximation_2}
&&\int_{t_0}^t
e^{-\int_{\tau}^{t}\beta_i(x_0(\chi),\tau_{i},\chi)d\chi}\phivec_i(x_0(\tau),p_{i},\tau)d\tau\simeq\\
&& \ \int_{t-T}^t
e^{-\int_{\tau}^{t}\beta_i(x_0(\chi),\tau_{i},\chi)d\chi}\phivec_i(x_0(\tau),p_{i},\tau)d\tau\triangleq\bar{\mu}_i(t,\tau_i,p_i),\nonumber
\end{eqnarray}
where $T\in\Real_{>0}$ is sufficiently large and $t\geq T+t_0$.

Alternatively, if $\phivec_i(x_0(t),p_{i},t)$,
$\beta_i(x_0(t),\tau_{i},t)$ are periodic with rationally -
dependent periods and satisfy the Dini condition in $t$,
integrals (\ref{eq:filtered}) can be estimated invoking a
Fourier expansion.
%
%
Notice that for continuous and Lipschitz in $p_i$ functions
$\mu_i(t,\tau_i,p_i)$ the coefficients of their Fourier
expansion remain continuous and Lipschitz with respect to
$p_i$.

In the next sections we present the general structure of the
observer for (\ref{eq:neural_model}) and provide a list of its
asymptotic properties.

\subsection{Observer definition and assumptions}

Consider the following function
$\varphivec(x_0,\lambdavec,t):\Real\times\Real^{s}\times\Real_{\geq0}\rightarrow\Real^{d}$,
$d=\sum_{i=0}^n d_i$:
\begin{equation}\label{eq:linear_regressor}
\begin{split}
 \varphivec(x_0,\lambdavec,t)&= \left( \phivec_0(x_0,p_0,t), c_1(x_0,q_1,t)\mu_1(t,\tau_1,p_1), \dots\right.\\
&\left. \dots, c_n(x_0,q_n,t)\mu_n(t,\tau_n,p_n)\right)^{T}
\end{split}
\end{equation}
The function $\varphivec(x_0,\lambdavec,t)$ is a concatenation
of $\phivec_0(\cdot)$ and integrals (\ref{eq:filtered}). We
assume that the values of $\varphivec(x_0,\lambdavec,t)$  can
be efficiently estimated for all $x_0$, $\lambdavec$, $t\geq 0$
up
 to a small mismatch. In other words, we suppose that there exists a
function  $\bar{\varphivec}(x_0,\lambdavec,t)$ such that the
following property
holds: 
\begin{equation}\label{eq:approximation_3}
\|\bar{\varphivec}(x_0,\lambdavec,t)-{\varphivec}(x_0,{\lambdavec},t)\|\leq
\Delta_\varphi, \ \Delta_\varphi\in\Real_{>0},
\end{equation}
where values of $\bar{\varphivec}(x_0,\lambdavec,t)$ are
efficiently computable for all $x_0$, $\lambdavec$, $t$  (see
e.g. (\ref{eq:approximation_2}) for an example of such
approximations), and $\Delta_{\varphi}$ is sufficiently small.

If parameters $\tau_i$, $p_i$, and  $q_i$ in the right-hand
side of (\ref{eq:neural_model}) would
 be known and $c_i(x_0,q_i,t)=1$, $\beta_i(x_0,\tau_i,t)=\tau_i$,
then the function ${\varphivec}(x_0,\lambdavec,t)$ could
 be estimated by $(\phivec_0(x_0,t),\eta_1,\dots,\eta_n)$ where $\eta_i$ are
 the solutions of the following auxiliary system (filter)
\begin{equation}\label{eq:auxiliary_filter}
\dot{\eta}_i=-\tau_i\eta_i + \phivec_i(x_0,p_i,t)
\end{equation}
with zero initial conditions. Systems like
(\ref{eq:auxiliary_filter}) are
 inherent components of standard adaptive observers \cite{Kreisselmeier:1977},
\cite{BastinGevers:1988}, \cite{Marino90}. In our case we
suppose that the values of $\tau_i$, $q_i$, $p_i$ are
 not know a-priori and that $c_i(x_0,q_i,t)$, $\beta_i(x_0,\tau_i,t)$ are not constant.
Therefore, we replace $\eta_i$ with their approximations, e.g.
as in (\ref{eq:approximation_2}):
\[
\begin{split}
 \bar{\varphivec}(x_0,\lambdavec,t)&= \left(\phivec_0(x_0,p_0,t), c_1(x_0,q_1,t)\bar{\mu}_1(t,\tau_1,p_1), \dots\right.\\
&\left. \dots,  c_n(x_0,q_n,t)\bar{\mu}_n(t,\tau_n,p_n)\right)^{T}.
\end{split}
\]
For periodic $\phivec_i(x_0(t),p_{i},t)$,
$\beta_i(x_0(t),\tau_{i},t)$ a Fourier expansion can be
employed to define $\bar{\varphivec}(x_0,\lambdavec,t)$. The
value of $\Delta_\varphi$ in (\ref{eq:approximation_3}) stands
for the accuracy of approximation, and as a rule of thumb the
more computational resources are devoted to approximate
$\varphivec(x_0,\lambdavec,t)$ the smaller is the value of
$\Delta_\varphi$.

With regard to the functions $\xi_i(t)$ in
(\ref{eq:neural_model}) we suppose that an upper bound,
$\Delta_\xi$, of the following sum is available:
\begin{equation}\label{eq:approximation_4}
\sum_{i=1}^n \frac{1}{\tau_i}\|\xi_i(\tau)\|_{\infty,[t_0,\infty]}+\|\xi_0(\tau)\|_{\infty,[t_0,\infty]}\leq \Delta_\xi, \ \Delta_{\xi}\in\Real_{\geq 0}.
\end{equation}

Denoting $c_0(x_0,q_0,t)=c_0(x_0,\lambdavec,t)$, for notational
convenience,
 we can now define the observer as
\begin{equation}\label{eq:linear_par_observer}
\left\{
\begin{split}
\dot{\hat{x}}_0&=-\alpha (\hat{x}_0-x_0) + \hat{\thetavec}^{T}\bar{\varphivec}(x_0,\hat{\lambdavec},t)+c_0(x_0,\hat{\lambdavec},t)+u(t)\\
\dot{\hat{\thetavec}}&=-\gamma_\theta
(\hat{x}_0-x_0)\bar{\varphivec}(x_0,\hat{\lambdavec},t), \
\gamma_\theta,\alpha\in\Real_{>0}
\end{split}\right.
\end{equation}
\begin{equation}\label{eq:linear_state_observer}
\dot{\hat{x}}_i= - \beta_i(x_0,\hat{\tau}_i,t) {\hat{x}}_i+ \hat{\theta}_i^{T} \phivec_i(x_0,\hat{p}_i,t), \ i=\{1,\dots,n\},
\end{equation}
where
$$\hat{\thetavec}=\col(\hat{\theta}_0,\hat{\theta}_1,\cdots,\hat{\theta}_n)$$
is the vector of estimates of $\thetavec$. The components of
vector
$\hat{\lambdavec}=\col(\hat{p}_0,\hat{q}_0,\hat{\tau}_1,\hat{p}_1,\hat{q}_1,\dots,\hat{\tau}_n,\hat{p}_n,\hat{q}_n)=\col(\hat{\lambda}_1,\dots,\hat{\lambda}_s)$,
with $s=\dim{(\lambdavec)}$, evolve according to the following
equations
\begin{equation}\label{eq:nonlinear_par_observer}
\begin{split}
&\left\{\begin{array}{ll}
\dot{\hat{x}}_{1,j}&=\gamma\cdot\omega_j \cdot e \cdot \left(\hat{x}_{1,j} - \hat{x}_{2,j} - \hat{x}_{1,j}\left(\hat{x}_{1,j}^2+\hat{x}_{2,j}^2\right)\right)\\
\dot{\hat{x}}_{2,j}&=\gamma\cdot\omega_j\cdot e \cdot \left(\hat{x}_{1,j}+ \hat{x}_{2,j} - \hat{x}_{2,j}\left(\hat{x}_{1,j}^2+\hat{x}_{2,j}^2\right) \right)\\
\hat{\lambda}_j(\hat{x}_{1,j})&=\lambda_{j,\min}+\frac{\lambda_{j,\max}-\lambda_{j,\min}}{2}(\hat{x}_{1,j}+1),
\\
e&=\sigma(\|x_0-\hat{x}_0\|_\varepsilon),\end{array}\right.
\end{split}
\end{equation}
\begin{equation}\label{eq:initial_conditions}
j=\{1,\dots,s\}, \ \hat{x}_{1,j}^2(t_0)+\hat{x}_{2,j}^2(t_0)=1,
\end{equation}
where $\sigma(\cdot):\Real\rightarrow\Real_{\geq 0}$ is a
bounded continuous function, i.e. $\sigma(\upsilon)\leq
S\in\Real_{>0}$, and $|\sigma(\upsilon)|\leq |\upsilon|$ for
all $\upsilon\in\Real$. We set $\omega_j\in\Real_{>0}$ and let
$\omega_j$ be {\it rationally-independent}:
\begin{equation}\label{eq:rational_independence}
\sum \omega_j k_j\neq 0, \ \forall \ k_j\in \Numbers.
\end{equation}

In order to proceed further we will need the notions of {\it
$\lambda$-uniform persistency of excitation} \cite{Lorea_2002}
and {\it nonlinear persistency of excitation} \cite{Cao_2003}:

\vspace*{0.25cm}
\begin{definition}[$\lambda$-uniform persistency of excitation] Let
$\varphivec:\Real_{\geq 0}\times
\mathcal{D}\rightarrow\Real^{n\times m}$,
 $\mathcal{D}\subset\Real^s$ be a continuous function.
We say that  $\varphivec(t,\lambdavec)$ is $\lambda$-uniformly
 persistently exciting ($\lambda$-uPE) if there
 exist $\mu\in\Real_{>0}$, $L\in\Real_{>0}$ such that for each $\lambdavec\in\mathcal{D}$
\begin{equation}\label{eq:PE_linear_uniform}
\int_{t}^{t+L}\varphivec(t,\lambdavec)\varphivec(t,\lambdavec)^T
d\tau \geq  \mu I \ \forall t\geq t_0.
\end{equation}
\end{definition}
In contrast to conventional definitions, the present notion
requires that the lower bound for the integral
$\int_{t}^{t+L}\varphivec(t,\lambdavec)\varphivec(t,\lambdavec)^T
d\tau$ in  (\ref{eq:PE_linear_uniform}) does not vanish for all
$\lambdavec\in\mathcal{D}$, and is separated away from zero. We
need this property in order to determine the linear parts,
$\theta_i$,
 of the parametric uncertainties in model (\ref{eq:neural_model}).

To reconstruct the nonlinear part of the uncertainties,
$\lambdavec$,
 we will require that $\bar{\varphivec}(x_0,{\lambdavec},t)$ is nonlinearly
persistently exciting in $\lambdavec$. Here we adopt the
definition of
 nonlinear persistent excitation from \cite{Cao_2003} with
a minor modification. The modification is needed to account for
a possibility that
\[
\bar{\varphivec}(x_0,{\lambdavec},t)=\bar{\varphivec}(x_0,{\lambdavec}',t), \ \lambdavec\neq\lambdavec', \ t\in\Real,
\]
which is the case, for example if
$\bar{\varphivec}(x_0,{\lambdavec},t)$ is periodic in
$\lambdavec$. The modified notion is presented in Definition
\ref{defi:nonlinear_pe} below.

\vspace*{0.25cm}
\begin{definition}[Nonlinear persistency of excitation]\label{defi:nonlinear_pe} The function\\ $\bar{\varphivec}(x_0,{\lambdavec},t)$ is nonlinearly persistently exciting if there exist $L, \beta\in\Real_{>0}$ such that for all $\lambdavec,\lambdavec'\in\Omega_{\lambda}$ and $t\in\Real$ there exists $t'\in[t-L,t]$ ensuring that the following inequality holds
\begin{equation}\label{eq:nonlinear_pe}
\|\bar{\varphivec}(x_0,{\lambdavec},t)-\bar{\varphivec}(x_0,{\lambdavec}',t')\|\geq \beta \cdot \ \dist(\mathcal{E}(\lambda),\lambdavec'),
\end{equation}
\begin{equation}\label{eq:equivalence_class}
\mathcal{E}(\lambdavec)=\{\lambdavec'\in\Omega_{\lambda}|\ \bar{\varphivec}(x_0,{\lambdavec}',t)=\bar{\varphivec}(x_0,{\lambdavec},t) \ \forall \ t\in\Real\}
\end{equation}
\end{definition}
The symbol $\mathcal{E}(\lambdavec)$ denotes the equivalence
class for $\lambdavec$, and
$\dist(\mathcal{E}(\lambda),\lambdavec')$ in
(\ref{eq:nonlinear_pe}) substitutes the Euclidian norm in the
\cite{Cao_2003} original definition. The nonlinear persistency
of excitation condition (\ref{eq:nonlinear_pe}) is very similar
to its linear counterpart (\ref{eq:PE_linear_uniform}). In fact
(\ref{eq:PE_linear_uniform}) can be written in the form of
inequality (\ref{eq:nonlinear_pe}), cf. \cite{Morgan_77}. For
further discussion of these notions, see  \cite{Cao_2003},
\cite{Lorea_2002}.

\subsection{Asymptotic properties of the observer}

The main results of this section are provided in Theorems
\ref{theorem:neural_identification} and
\ref{theorem:parameter_identification}. Theorem
\ref{theorem:neural_identification} establishes conditions for
state boundedness of the observer, and states its general
asymptotic properties. Theorem
\ref{theorem:parameter_identification} specifies a set of
conditions for the possibility of asymptotic reconstruction of
$\theta_i$, $\tau_i$, and $p_i$, up to their equivalence
classes and small mismatch due to errors.

Proofs of Theorems \ref{theorem:neural_identification},
\ref{theorem:parameter_identification} and other auxiliary
results can be found in \cite{Adaptive_Observers}.
\vspace*{0.25cm}
\begin{theorem}[Boundedness]\label{theorem:neural_identification} Let system (\ref{eq:neural_model}), (\ref{eq:linear_par_observer}) -- (\ref{eq:nonlinear_par_observer}) be given. Assume that function
$\bar{\varphivec}(x_0(t),\lambdavec,t)$ is $\lambda$-uniformly
persistently exciting, and the functions
$\bar{\varphivec}(x_0(t),\lambdavec,t)$,
$c_0(x_0(t),\lambdavec,t)$  are Lipschitz in $\lambdavec$:
\begin{equation}\label{eq:thm:1}
\begin{split}
&\|\bar{\varphivec}(x_0(t),\lambdavec,t)-\bar{\varphivec}(x_0(t),\lambdavec',t)\|\leq
D \|\lambdavec-\lambdavec'\|,\\
&\|c_0(x_0(t),\lambdavec,t)-c_0(x_0(t),\lambdavec',t)\|\leq
D_c \|\lambdavec-\lambdavec'\|.
\end{split}
\end{equation}
Then there exist numbers $\varepsilon>0$, $\gamma^\ast>0$ such
that for all $\gamma \in(0,\gamma^\ast]$:
\begin{enumerate}
\item [1)] trajectories of the closed loop system
    (\ref{eq:linear_par_observer}) --
    (\ref{eq:nonlinear_par_observer}) are bounded and
\begin{equation}\label{eq:asymptotic:output}
\lim_{t\rightarrow\infty} \|\hat{x}_0(t)-x_0(t)\|_\varepsilon=0;
\end{equation}

\item [2)] there exists
    $\lambdavec^\ast\in\Omega_{\lambda}$,
    $\kappa\in\Real_{>0}$ such that
\begin{equation}\label{eq:asymptotic:linear:parameter}
\begin{split}
\lim_{t\rightarrow\infty}\hat{\lambdavec}(t)&=\lambdavec^\ast\\
\limsup_{t\rightarrow\infty}\|\hat{\thetavec}(t)-\thetavec\|&<\kappa ((D \|\thetavec\| + D_c) \|\lambdavec^{\ast}-\lambdavec\|+2\Delta).
\end{split}
\end{equation}
\begin{equation}\label{eq:delta_definition}
\Delta=\|\thetavec\|\Delta_{\varphi}+\Delta_{\xi}
\end{equation}
\end{enumerate}
\end{theorem}

\vspace*{0.25cm}
\begin{remark}\normalfont
Theorem \ref{theorem:neural_identification} assures that the
estimates $\hat{\thetavec}(t)$, $\hat{\lambdavec}(t)$
asymptotically converge to a neighborhood of the actual values
$\thetavec$, $\lambdavec$. It does not specify, however, how
close these estimates are to the true values of $\thetavec$,
$\lambdavec$.

The next result states that if the values of $\Delta_\varphi$
and $\Delta_{\xi}$:
\[
\begin{split}
&\|\varphivec(x_0,\lambdavec,t)-\bar{\varphivec}(x_0,\lambdavec,t)\|\leq \Delta_{\varphi}\\
& \sum_{i=1}^n \frac{1}{\tau_i}\|\xi_i(\tau)\|_{\infty,[t_0,\infty]}+\|\xi_0(\tau)\|_{\infty,[t_0,\infty]}\leq \Delta_\xi
\end{split}
\]
in (\ref{eq:approximation_3}), (\ref{eq:approximation_4}) are
small,  e.g. $\bar{\varphivec}(x_0(t),\lambdavec,t)$
approximates ${\varphivec}(x_0(t),\lambdavec,t)$ with
sufficiently high accuracy and the unmodeled dynamic is
negligible, the estimates $\hat{\thetavec}(t)$,
$\hat{\lambdavec}(t)$ will converge to small neighborhoods of
the equivalence classes of $\thetavec$, $\lambdavec$. The sizes
of these neighborhoods are shown to be bounded from above by
monotone functions $\Delta$:
\[
\Delta=\|\thetavec\|\Delta_\varphi+\Delta_\xi
\]
vanishing at zero.
Formally this result is stated in Theorem
\ref{theorem:parameter_identification} below
\end{remark}
\vspace*{0.25cm}
\begin{theorem}[Convergence]\label{theorem:parameter_identification} Let the assumptions of Theorem \ref{theorem:neural_identification} hold, assume that $\bar{\varphi}_0(x_0,\lambdavec,t)\in\mathcal{C}^{1}$, the derivative $\pd \bar{\varphivec}_0(x_0(t),\lambdavec,t)/\pd t$ is globally bounded,
 and $\Delta=\|\thetavec\|\Delta_\varphi + \Delta_\xi$ is small. Then there exist numbers $\varepsilon>0$, $\gamma^\ast>0$ such that
for all $\gamma_i\in(0,\gamma^\ast)$
\begin{enumerate}
\item [1)]
\[
\limsup_{t\rightarrow\infty}\|\hat{\thetavec}(t)-\thetavec\|=\mathcal{O}(\sqrt{\Delta})+\mathcal{O}(\Delta)
\]

\item [2)] in case
    $\thetavec^{T}\bar{\varphivec}(x_0(t),\lambdavec,t)+c_0(x_0(t),\lambdavec,t)$
    is nonlinearly persistently exciting with respect to
    $\lambdavec$, then the estimates $\hat{\lambdavec}(t)$
    converge into a small vicinity of
    $\mathcal{E}(\lambda)$:
\begin{equation}\label{eq:asymptotic:nonlinear:parameter}
\limsup_{t\rightarrow\infty} \ \dist(\hat{\lambdavec}(t),\mathcal{E}(\lambdavec))=\mathcal{O}(\sqrt{\Delta})+\mathcal{O}(\Delta)
\end{equation}
\end{enumerate}
\end{theorem}

\vspace*{0.5cm}
\setcounter{equation}{0}
\section{Examples}\label{sec:examples}

\subsection{Parameter estimation of the 2D Hindmarsh-Rose model with Bastin-Gevers observer}

\begin{figure}
 \centering
 \subfigure
[$\hat{\eta}_1(t)$ v. t]
 {\resizebox{5cm}{5cm}{\includegraphics{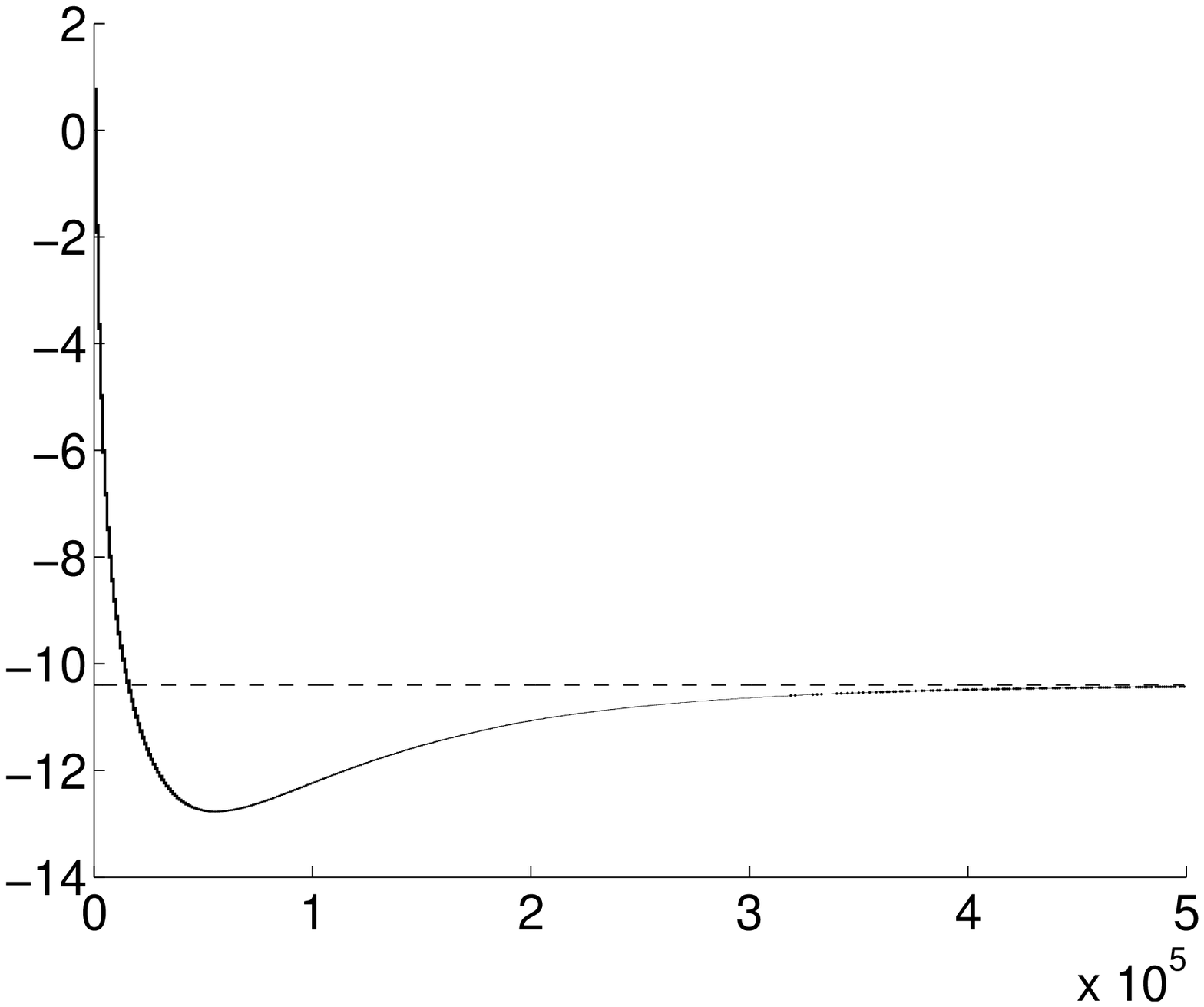}}}
 \subfigure
[$\hat{\eta}_2(t)$ v. t]
 {\resizebox{5cm}{5cm}{\includegraphics{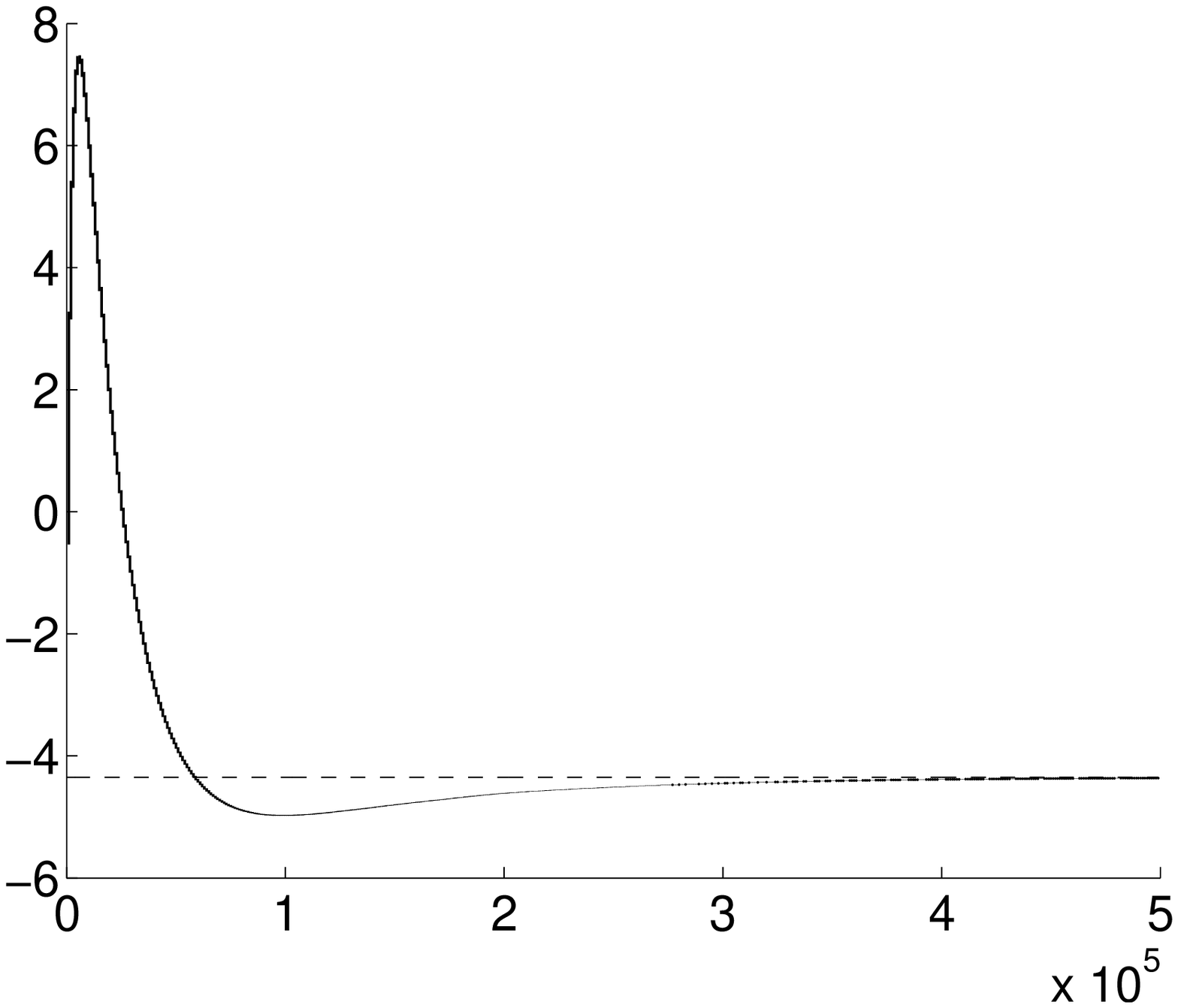}}}
 \subfigure
[$\hat{\eta}_3(t)$ v. t]
 {\resizebox{5cm}{5cm}{\includegraphics{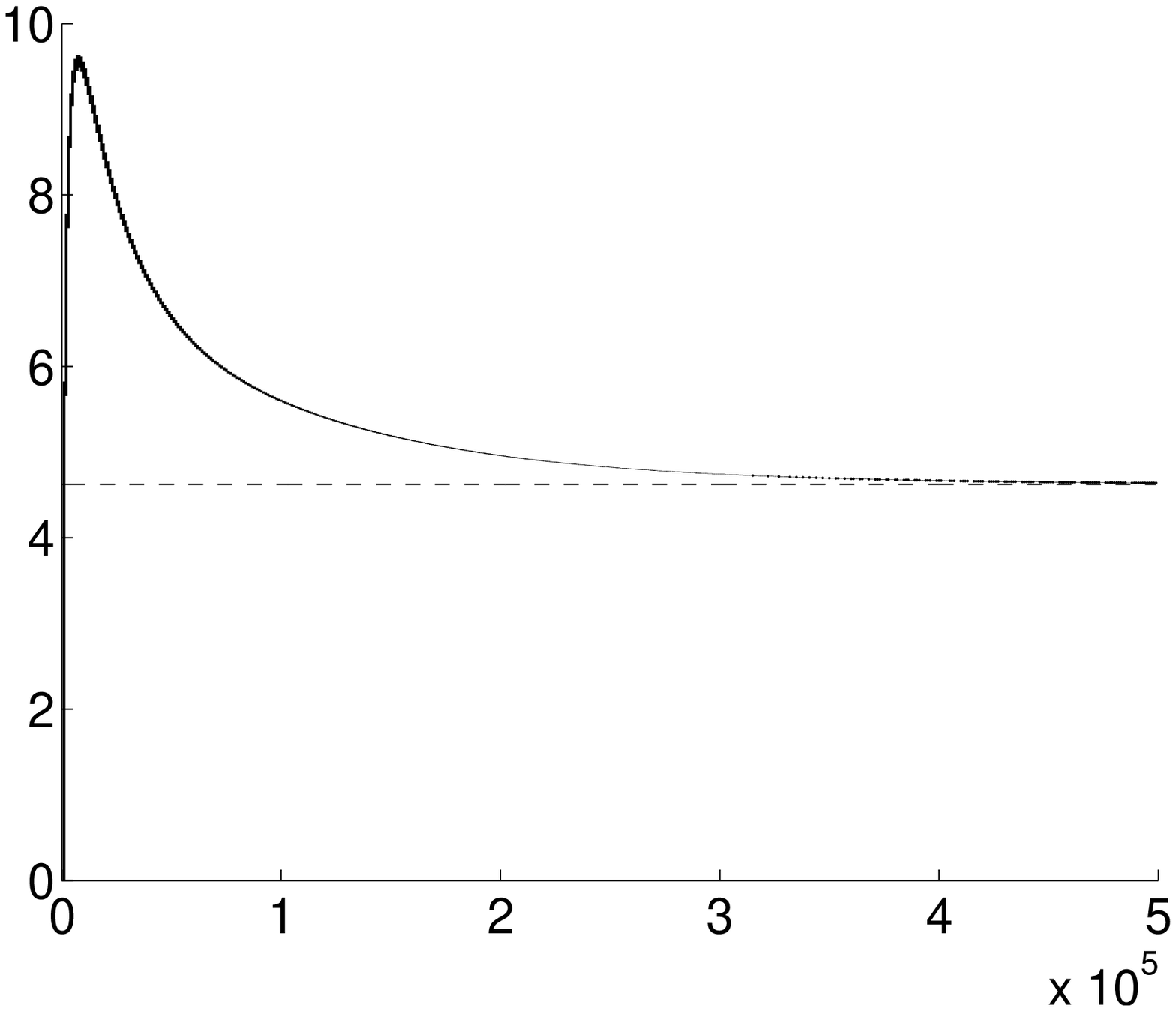}}}
 \subfigure
[$\hat{\eta}_5(t)$ v. t]
 {\resizebox{5cm}{5cm}{\includegraphics{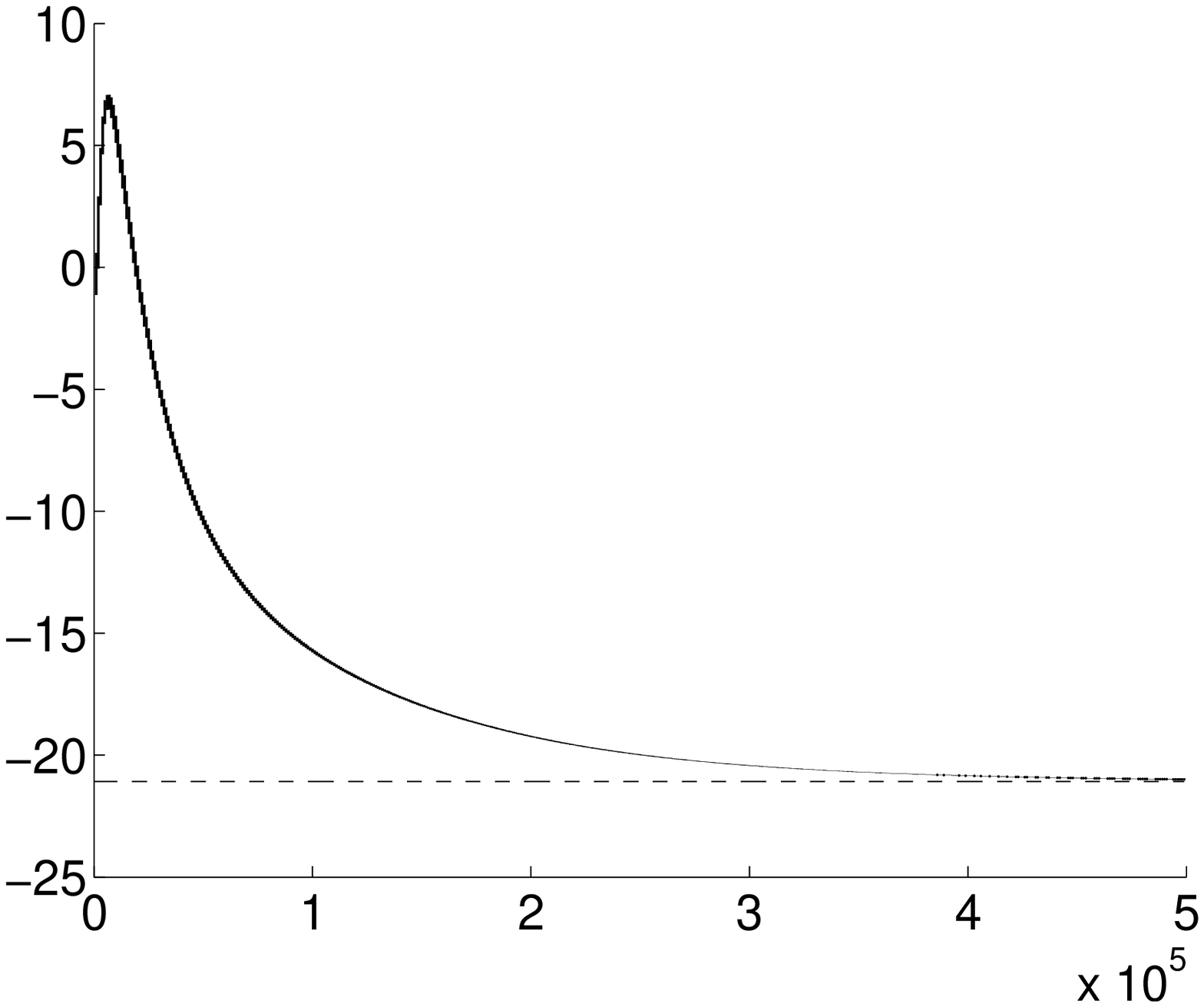}}}
 \subfigure
[$\hat{\eta}_6(t)$ v. t]
 {\resizebox{5cm}{5cm}{\includegraphics{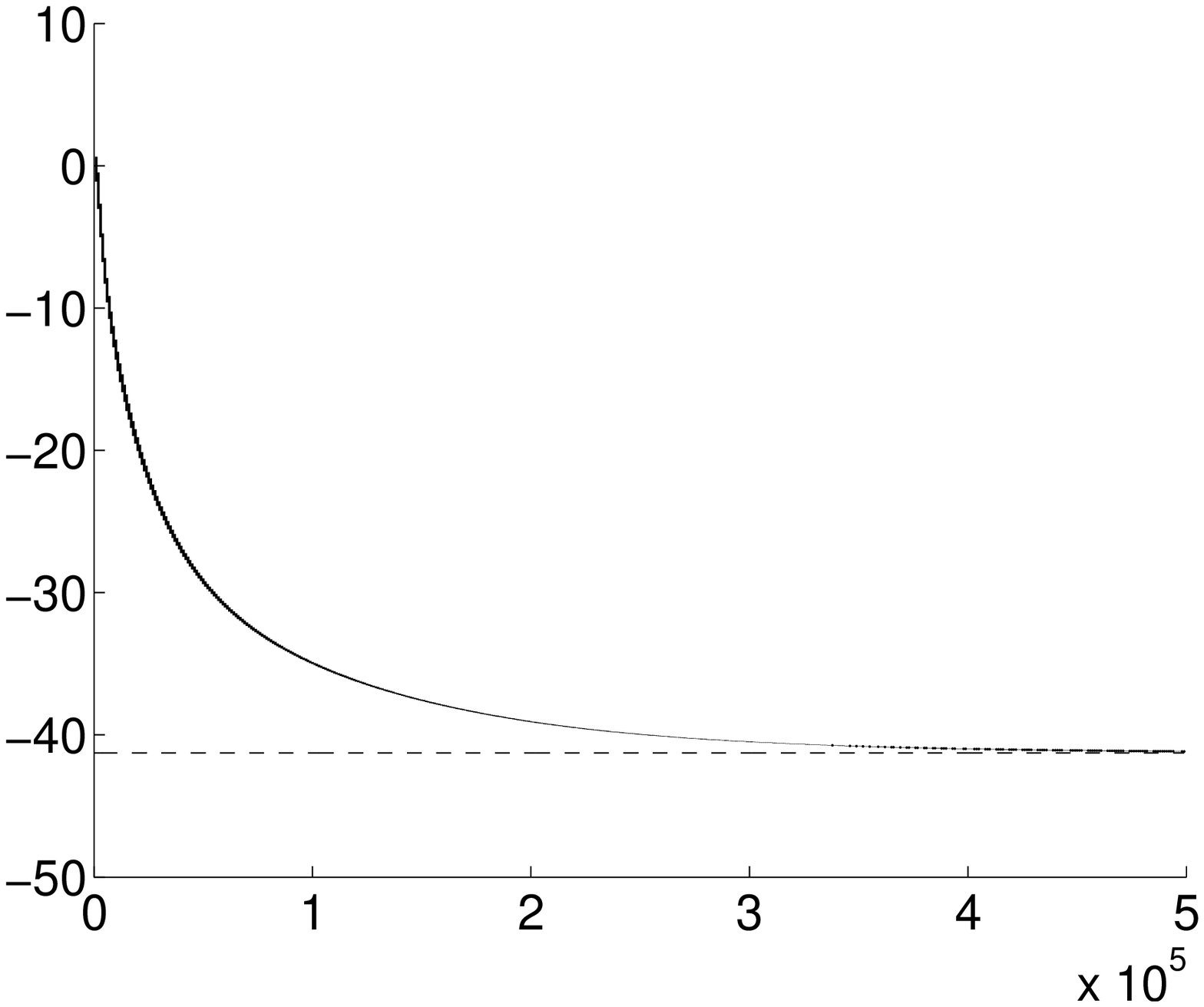}}}
 \subfigure
[$\hat{\eta}_7(t)$ v. t]
 {\resizebox{5cm}{5cm}{\includegraphics{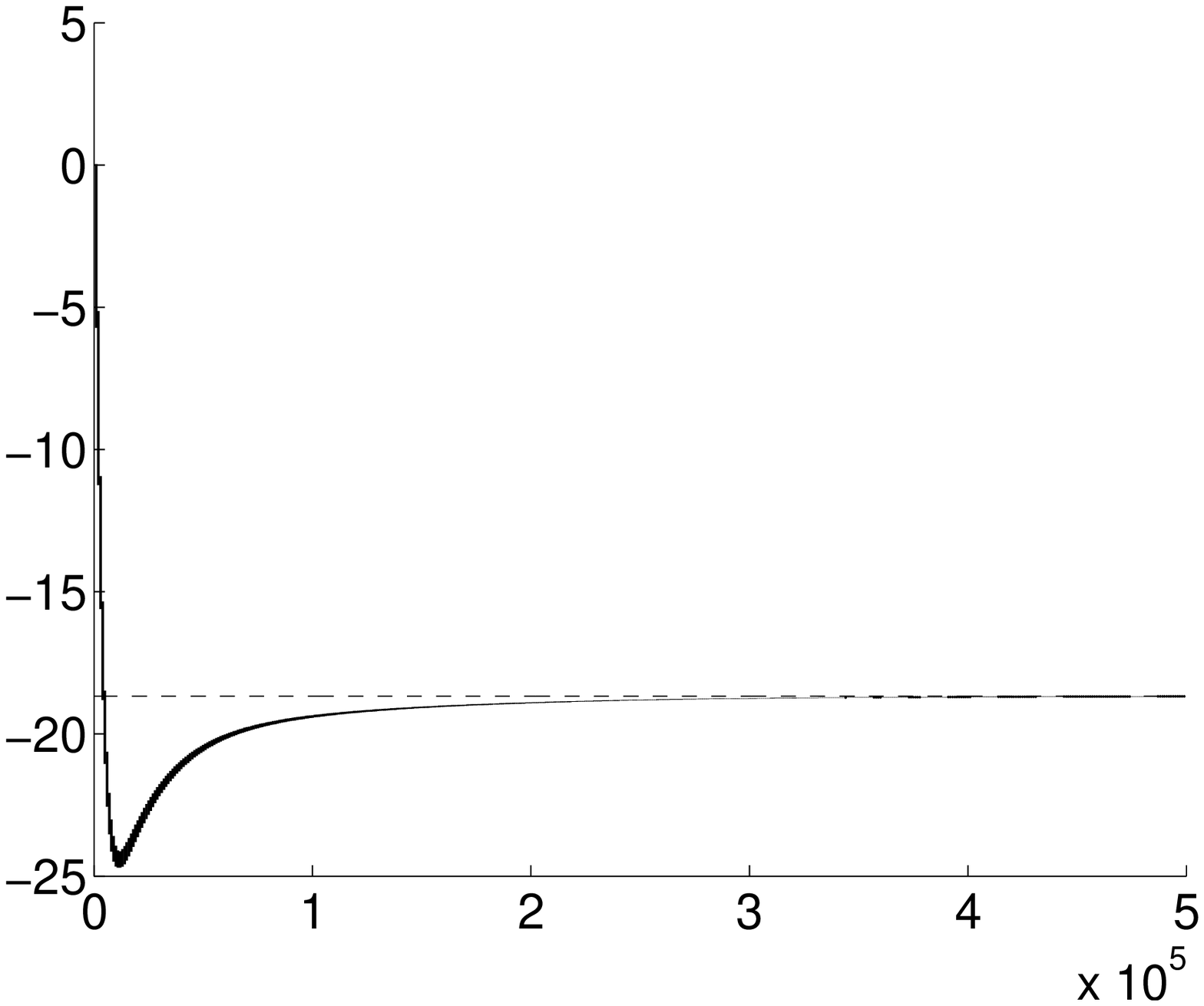}}}
 \subfigure
[$\hat{\eta}_8(t)$ v. t]
 {\resizebox{5cm}{5cm}{\includegraphics{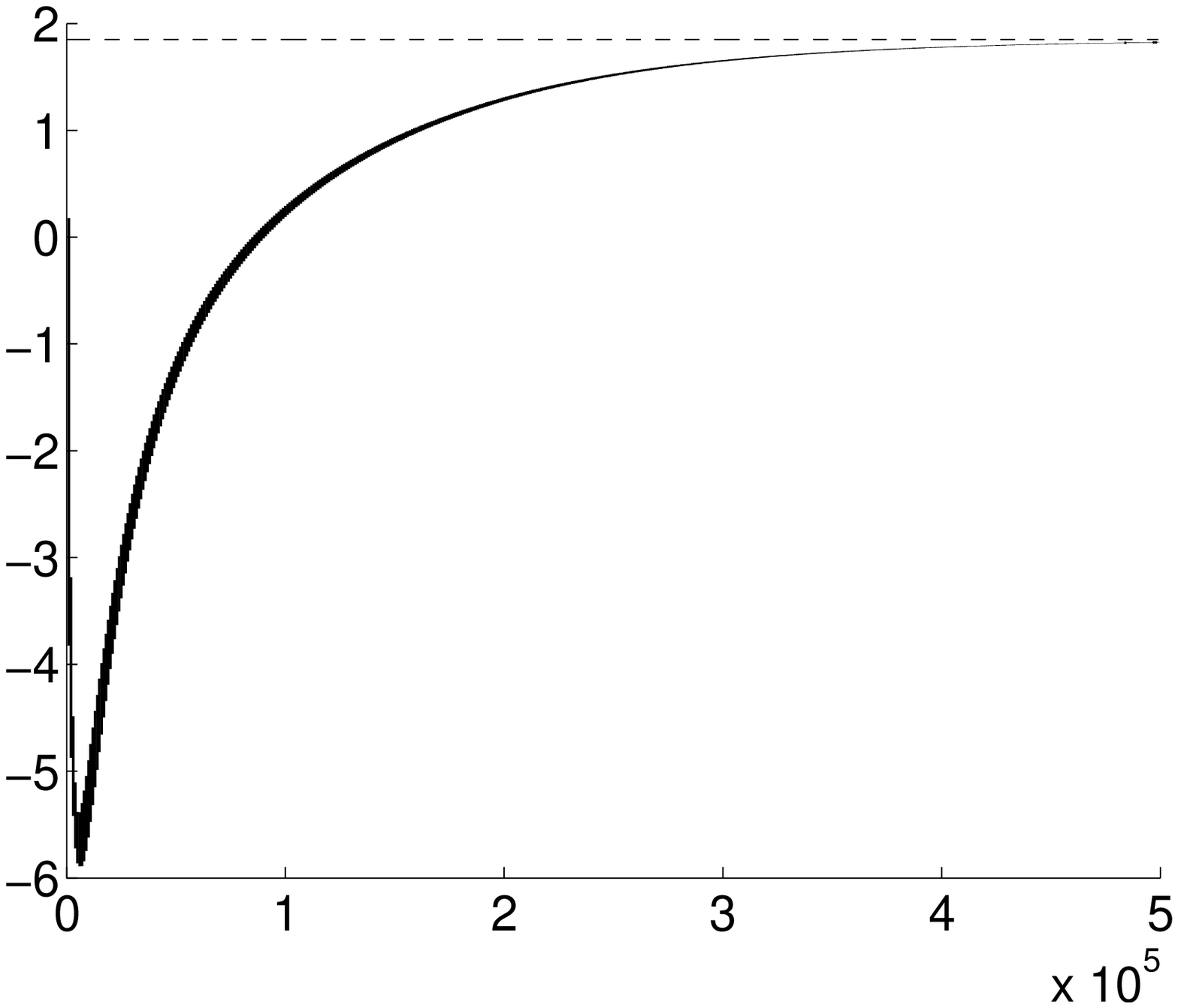}}}
\caption{Simulation results for Bastin-Gevers observer. Each $\hat{\eta}_i(t)$, $i\ne 4$ are shown
with the true values indicated with a broken line.
The periodic time of the spikes is circa $10$ seconds while the total time simulated is $5\times 10^5$ seconds,
thus the figure spans thousands of spike cycles.
 The visual effect of these extreme time scales is seen in the figure;
 hunting oscillations within the observer are seen
as thickening of the graphs of certain estimated parameters.
}
\label{fig:BG_simulation}
\end{figure}

The canonical form (\ref{eqn_BG1}) and the observer presented
in Section 3.3 were built in MATLAB and using
the differential equation solver ode45, numerical results were
obtained. Figure \ref{fig:BG_simulation} shows the parameter
convergence of each $\hat{\eta}_i(t)$, $i\ne 4$. The various
parameter values were set as follows: for the neuron

\[
 \lambda = 2.027,
 \theta_{13} = -10.4,
 \theta_{12} = -4.35,
 \theta_{11} = 6.65,
 \theta_{10} = 0.9125,
 \theta_{22} = -32.45,
 \theta_{11} = -32.15
\]

and for the observer dynamics

\[
 k = 1, c_1 = 1, \Gamma = \mathrm{diag}(1,1,1,1,1,1,1,1), F = -1
\]

\subsection{Parameter estimation of the 2D Hindmarsh-Rose model with Mario-Tomei observer}

\begin{figure}
 \centering
 \subfigure
[$\hat{\upsilon}_1(t)$ v. t]
 {\resizebox{5cm}{5cm}{\includegraphics{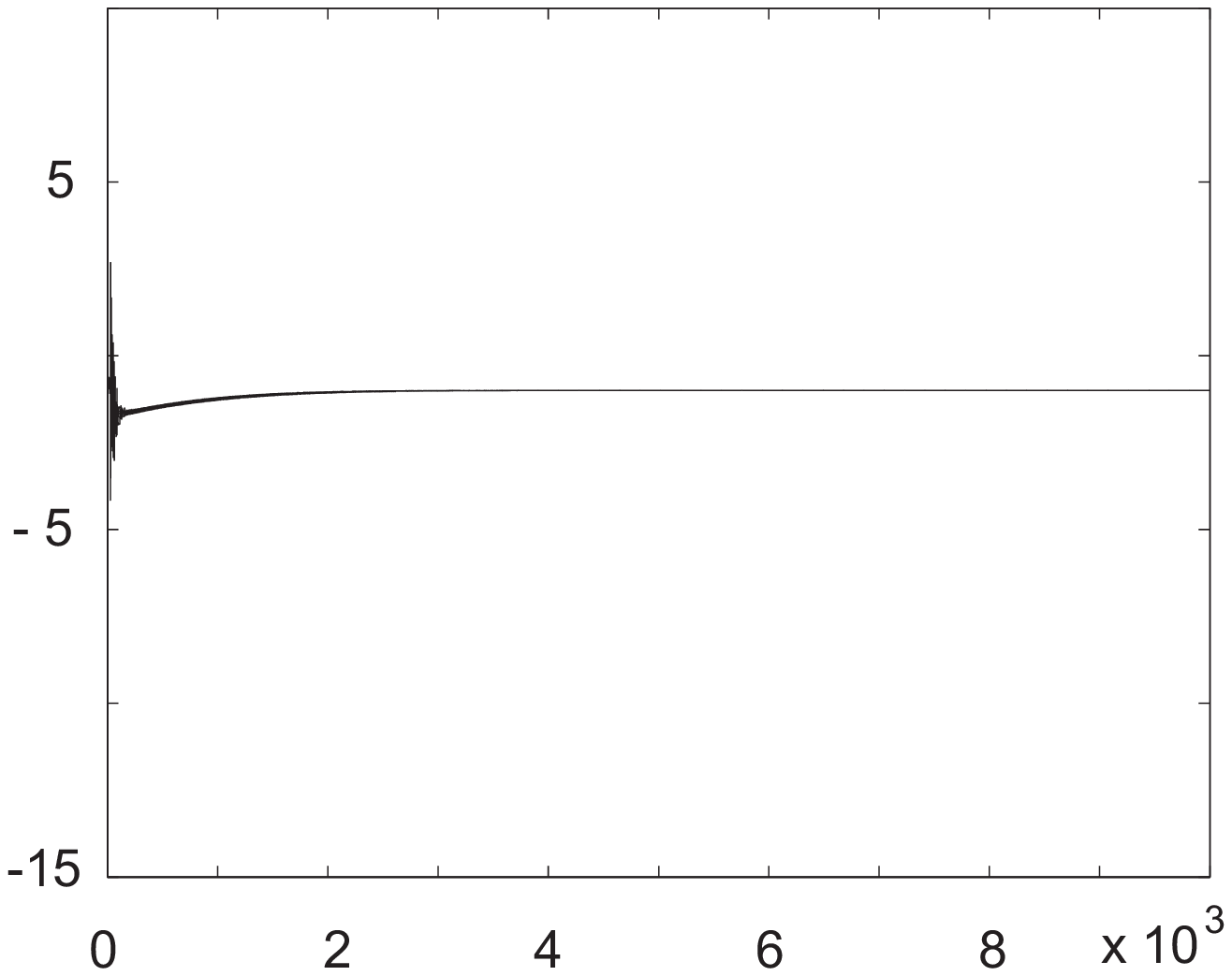}}}
 \subfigure
[$\hat{\upsilon}_2(t)$ v. t]
 {\resizebox{5cm}{5cm}{\includegraphics{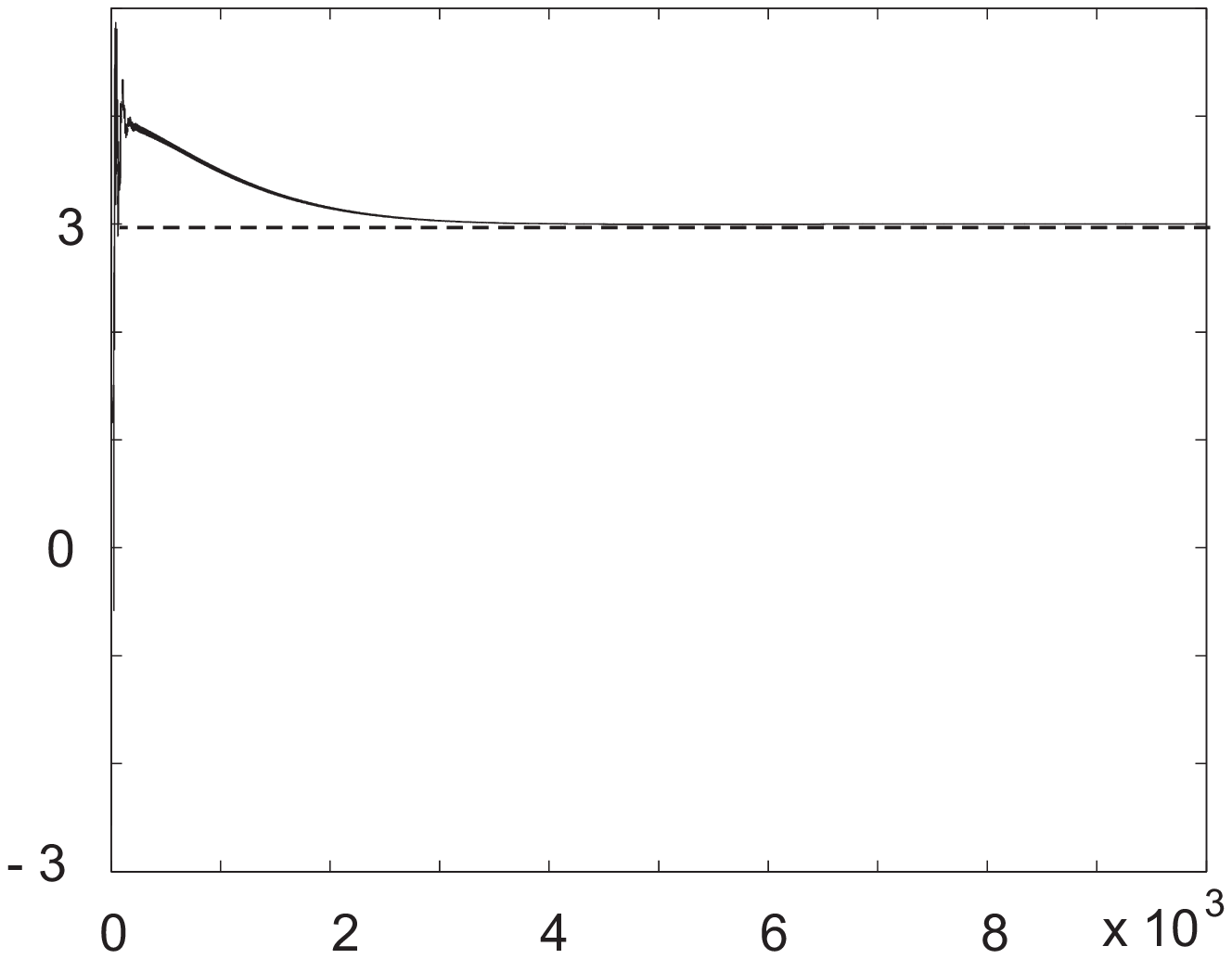}}}
 \subfigure
[$\hat{\upsilon}_3(t)$ v. t]
 {\resizebox{5cm}{5cm}{\includegraphics{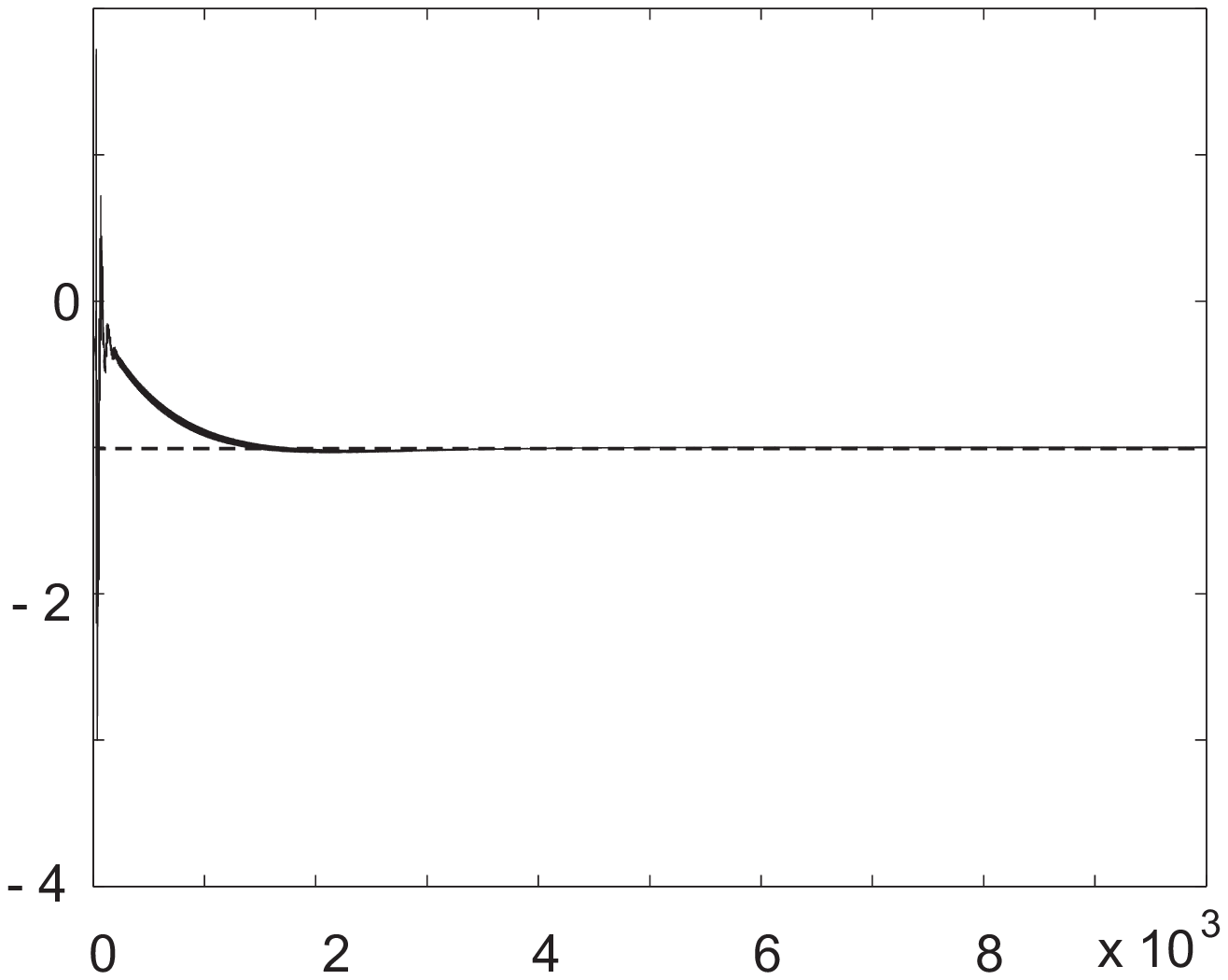}}}
 \subfigure
[$\hat{\upsilon}_4(t)$ v. t]
 {\resizebox{5cm}{5cm}{\includegraphics{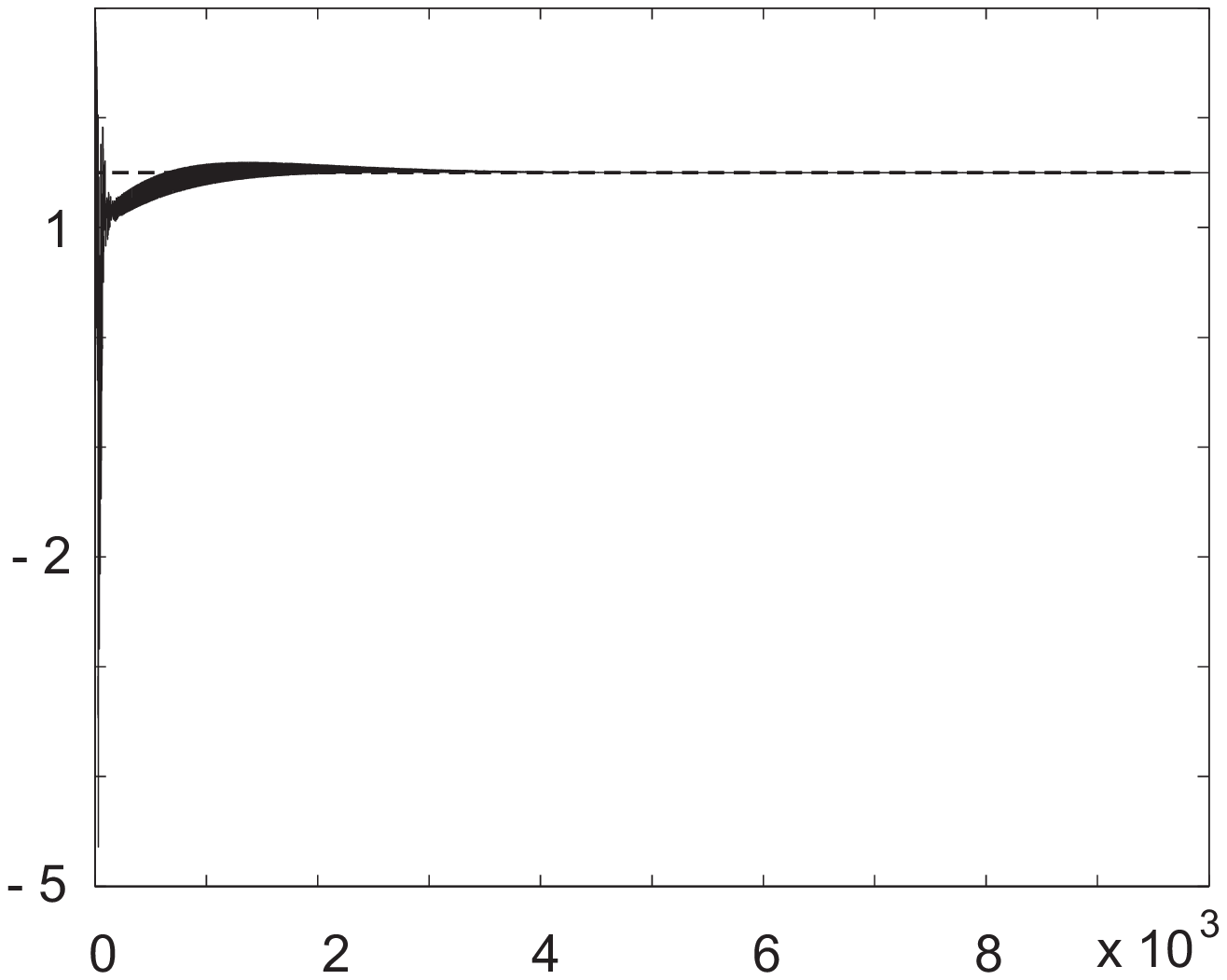}}}
 \subfigure
[$\hat{\upsilon}_5(t)$ v. t]
 {\resizebox{5cm}{5cm}{\includegraphics{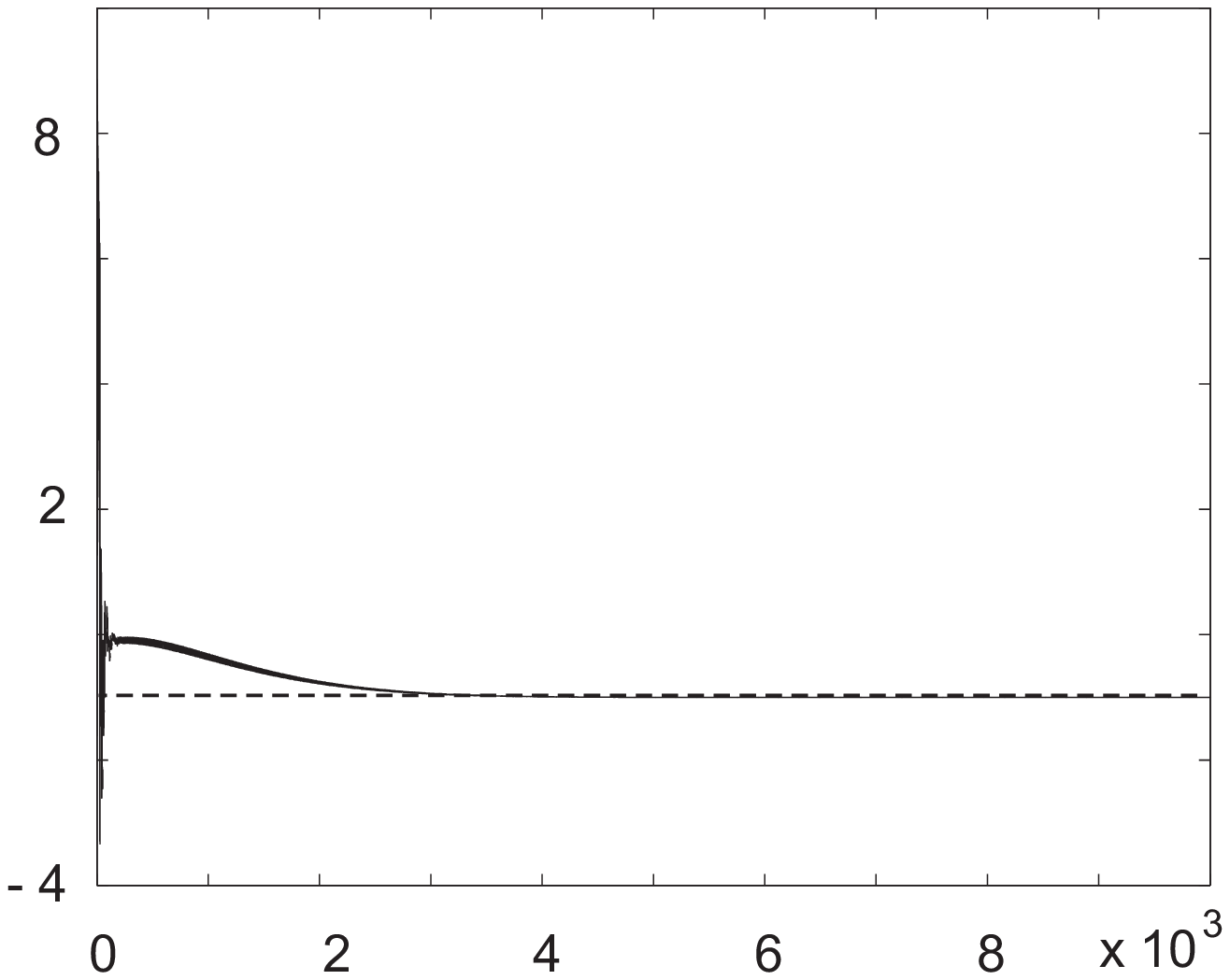}}}
 \subfigure
[$\hat{\upsilon}_6(t)$ v. t]
 {\resizebox{5cm}{5cm}{\includegraphics{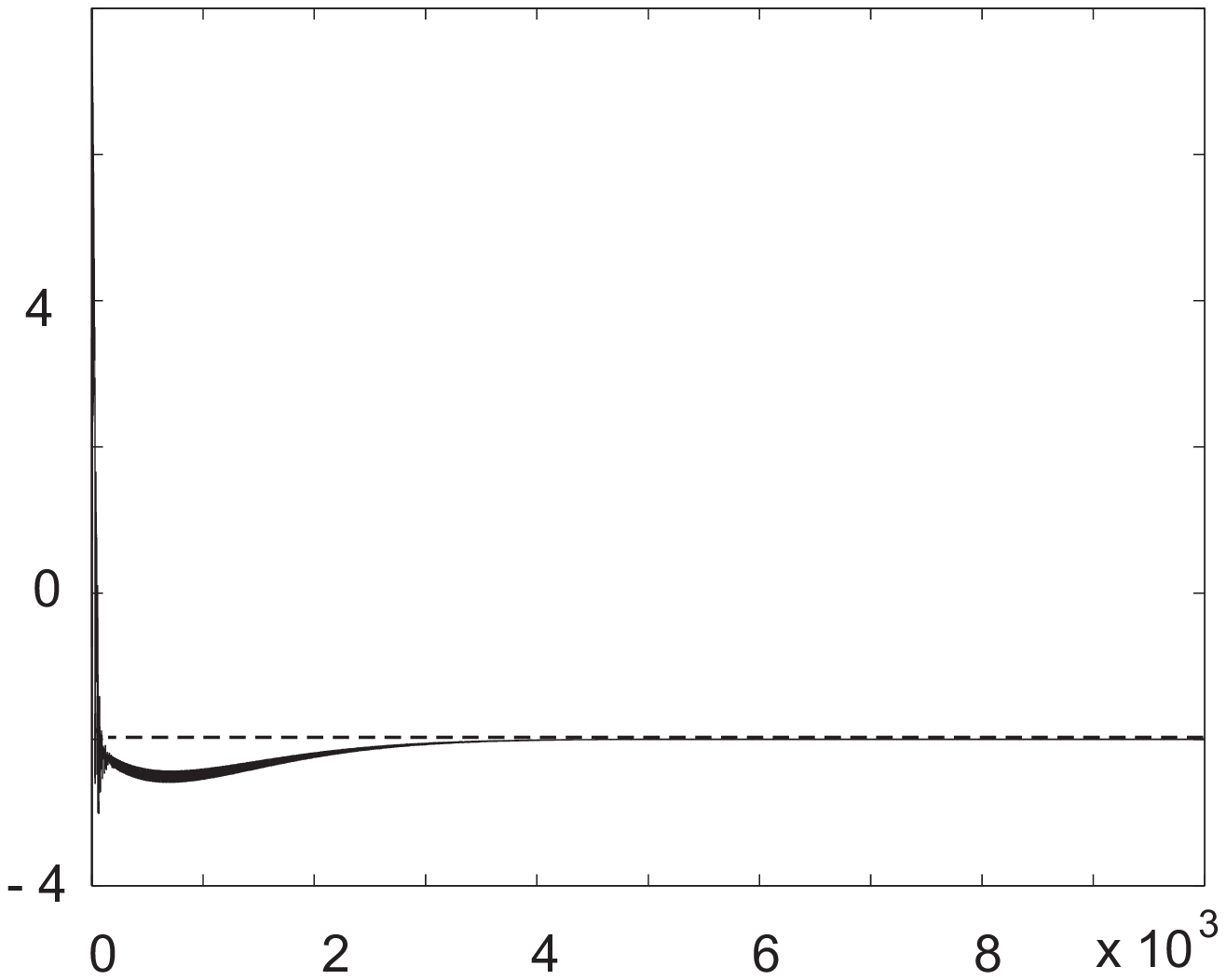}}}
 \subfigure
[$\hat{\upsilon}_7(t)$ v. t]
 {\resizebox{5cm}{5cm}{\includegraphics{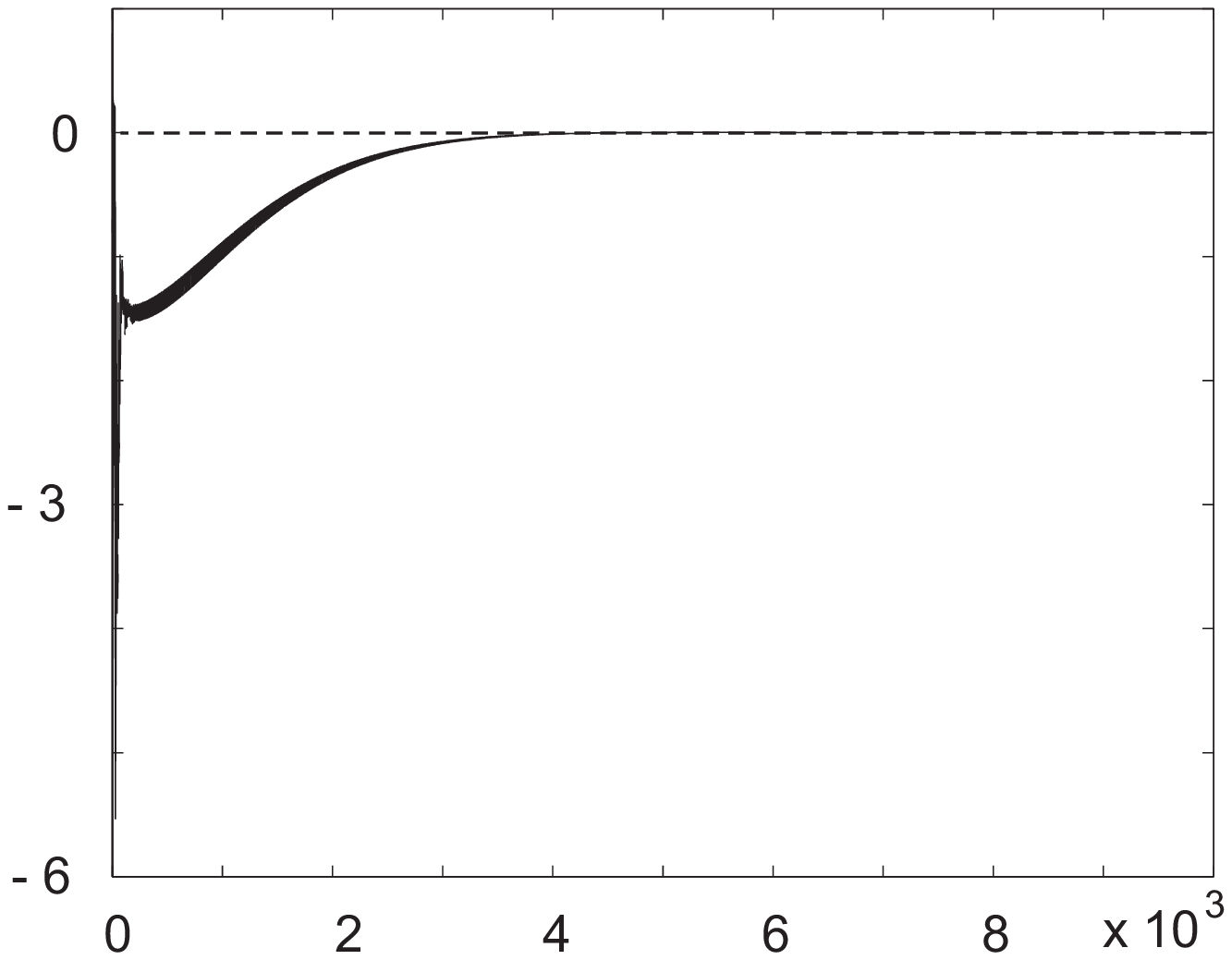}}}
\caption{Simulation results for Marino-Tomei observer.
Each $\hat{\upsilon}_i(t)$ are shown with the true values indicated with a broken line.
The periodic time of the spikes is circa $10$ seconds while the total time simulated is $10\times 10^3$ seconds,
thus the figure spans thousands of spike cycles.
 The visual effect of these extreme time scales is seen in the figure;
 hunting oscillations within the observer are seen
as thickening of the graphs of certain estimated parameters.
}
\label{fig:MT_simulation}
\end{figure}

In addition to simulating the Bastin-Gever observer for the
Hindmarsh-Rose model we also simulated
 observer (\ref{eq:passivity_observer}) derived within the framework of the approach presented
in \cite{Marino90}. In this case we set
\[
\theta_{1,3}=-1, \ \theta_{1,2}=3, \ \theta_{1,3}=0, \ \theta_{1,0}= 1.5, \ \theta_{2,3}=0, \ \theta_{2,2}=-5, \ \theta_{2,1}=0, \ \lambda=-1
\]
There is no particular reasoning behind our choice of
parameters in both this and previous case apart from that these
parameters must induce persistent oscillatory dynamics of the
solutions
 of (\ref{eq:HR:2}). Parameters  of the observer were chosen as follows:
\[
l_1=l_2=1, \ \gamma=1, \ k=1.
\]
Simulation results for this system are shown in Figure
\ref{fig:MT_simulation}.

\subsection{Parameter estimation of the Morris-Lecar model}

Let us now turn to a more realistic class of equations, i.e.
conductance-based models. In particular, we consider the
Morris-Lecar model, \cite{Morris_Lecar}:
\begin{equation}\label{eq:Morris_Lecar}
\begin{split}
\dot{V}&=\frac{1}{C}\left(-\bar{g}_{\mathrm{Ca}}m_{\infty}(V)(V-E_{\mathrm{Ca}})-\bar{g}_{K}w(V-E_{K})-\bar{g}_{L}(V-V_{0})\right)+I\\
\dot{w}&=-\frac{1}{\tau(V)}w + \frac{w_\infty(V)}{\tau(V)}
\end{split}
\end{equation}
where
\[
\begin{split}
m_{\infty}(V)&=0.5\left(1+\tanh\left(\frac{V-V_1}{V_2}\right)\right)\\
w_{\infty}(V)&=0.5\left(1+\tanh\left(\frac{V-V_3}{V_4}\right)\right)\\
\tau(V)&=T_0 \frac{1}{\cosh\left(\frac{V-V_{3}}{2V_{4}}\right)}
\end{split}
\]
System (\ref{eq:Morris_Lecar}) is a reduction of the standard
$4$-dimensional Hodgkin-Huxley equations, and is one of the
simplest models describing the dynamics of evoked membrane
potential and, at the same time, claiming biological
plausibility.

Parameters $\bar{g}_{\mathrm{Ca}}$, $\bar{g}_{K}$, and
$\bar{g}_{L}$ stand for the maximal conductances of the
calcium,  potassium and leakage currents respectively; $C$ is
the membrane capacitance; $V_1$, $V_2$, $V_3$, $V_4$ are the
parameters of the gating variables; $T_0$ is the parameter
regulating the time scale of ionic currents; $E_{\mathrm{Ca}}$
and $E_{\mathrm{K}}$ are the Nernst potentials of the calcium
and potassium currents, and $E_L$ is the rest potential.
Variable $I$ models an external stimulation current. In this
example the value of $I$ was set to $I=10$.

The total number of parameters in system
(\ref{eq:Morris_Lecar}) is $12$, excluding the stimulation
current $I$. Some of these parameters, however, are already
available or can be considered typical. For example the values
of the Nernst potentials for calcium and potassium channels,
$E_{\mathrm{Ca}}$, $E_{\mathrm{K}}$, are known and usually are
set as follows $E_{\mathrm{Ca}}=100$, $E_{\mathrm{K}}=-70$. The
value of the rest potential, $V_{0}$, can be estimated from the
cell explicitly. Here we set $V_{0}=-50$. Parameters $V_1$,
$V_2$ characterize the steady-state response curve of the
activation gates corresponding to the calcium channels, and
$V_3$, $V_4$ are the parameters of the potassium channels. In
the simulations we set these parameters to standard values as
e.g. in \cite{Koch_2002}: $V_1=-1$, $V_2=15$, $V_3=10$, and
$V_4=29$.

The values of parameters, $\bar{g}_{\mathrm{Ca}}$,
$\bar{g}_{K}$, $\bar{g}_{L}$, and $T_0$,  however, may vary
substantially from one cell to another. For example, the values
of $\bar{g}_{\mathrm{Ca}}$, $\bar{g}_{K}$, $\bar{g}_{L}$ depend
on the density of ion channels in a patch of the membrane; the
value of $T_0$ is dependent on temperature. Hence, in order to
model the dynamics of individual cells, we need to be able to
recover these values from data.

As before, we suppose that the values of  $V$ over time are
available for direct observation, and the values of $w$ are not
measured. System (\ref{eq:Morris_Lecar}) has no linear
time-invariant part, and the dynamics of $w$ are governed by a
nonlinear differential equation with the time-varying
relaxation factor, $1/\tau(V)$. Therefore, observers presented
in Section 3 may not be applied
explicitly to this system. This does not imply, however, that
parameters $\bar{g}_{\mathrm{Ca}}$, $\bar{g}_{K}$,
$\bar{g}_{L}$, and $T_0$ cannot be recovered from the
measurements of $V$. In fact, as we show below, one can
successfully reconstruct these parameters by using observers
defined in Section 4.

For the sake of notational consistency we denote $x_0=V$,
$x_1=w$, and without loss of generality suppose that $C=1$.
Hence system (\ref{eq:Morris_Lecar}) can now be rewritten as
follows:
\begin{equation}\label{eq:Morris_Lecar:1}
\begin{split}
\dot{x}_0&=\theta_{0,1} m_{\infty}(x_0)(x_0-E_{\mathrm{Ca}})+\theta_{0,1} x_1 (x_0-E_{K})+\theta_{0,3}(x_0-V_{0})+I\\
\dot{x}_1&=-\beta_1(x_0,\lambda)x_1 + \frac{1}{\lambda} \phi_1(x_0)
\end{split}
\end{equation}
where
\[
\begin{split}
\beta_1(x_0,\lambda)&=\frac{1}{\lambda} \cosh\left(\frac{x_0-V_{3}}{2V_{4}}\right), \ \lambda=T_0\\
\phi_1(x_0)&= \cosh\left(\frac{x_0-V_{3}}{2V_{4}}\right)w_\infty (x_0)
\end{split}
\]
Noticing that $\beta_1(x_0,\lambda)$ is separated away from
zero for all bounded $x_0$ and positive $\lambda$ we substitute
variable $x_1$ in (\ref{eq:Morris_Lecar:1}) with its estimation
\[
\chi(\lambda,t)=\int_{t-T}^{t} \frac{1}{\lambda} e^{-\frac{1}{\lambda} \int_{\tau}^{t} \cosh\left(\frac{x_0(s)-V_{3}}{2V_{4}}\right) ds}\cosh\left(\frac{x_0(\tau)-V_{3}}{2V_{4}}\right)w_\infty (x_0(\tau))d\tau
\]
The larger the value of $T$ the higher the accuracy of
estimation for large $t$. After this substitution system
(\ref{eq:Morris_Lecar:1}) reduces to just only one equation
\begin{equation}\label{eq:Morris_Lecar:2}
\begin{split}
\dot{x}_0&=\theta_{0,1} \phi_{0,1}(x_0) + \theta_{0,2} \phi_{0,2}(x_0,\lambda,t) + \theta_{0,3} \phi_{0,3}(x_0)+I + \xi_0(t)
\end{split}
\end{equation}
where
\[
\begin{split}
\phi_{0,1}(x_0)&=m_{\infty}(x_0)(x_0-E_{\mathrm{Ca}})\\
\phi_{0,2}(x_0,\lambda,t)&=(x_0-E_{K})\chi(\lambda,t)\\
\phi_{0,3}(x_0)&=(x_0-V_{0})
\end{split}
\]
and term $\xi_0(t)$ is bounded.

Equation (\ref{eq:Morris_Lecar:2}) is a special case of
(\ref{eq:neural_model}), and hence we can apply the results of
Section 4 to construct an observer for
asymptotic estimation of the values of $\theta_{0,1}$,
$\theta_{0,2}$, $\theta_{0,3}$, and $\lambda$. In accordance
with  (\ref{eq:linear_par_observer}) --
(\ref{eq:nonlinear_par_observer}) we obtain the following
observer equations
\begin{equation}\label{eq:Morris_Lecar:linear}
\begin{split}
\dot{\hat{x}}_0&=-\alpha (\hat{x}_0-x_0) +  \hat{\theta}_{1} \phi_{0,1}(x_0) + \hat{\theta}_{2} \phi_{0,2}(x_0,\hat{\lambda},t) + \hat{\theta}_{3} \phi_{0,3}(x_0)+I\\
\dot{\hat{\theta}}_1&=-\gamma_\theta (\hat{x}_0-x_0)\phi_{0,1}(x_0)\\
\dot{\hat{\theta}}_2&=-\gamma_\theta (\hat{x}_0-x_0)\phi_{0,2}(x_0,\hat{\lambda},t)\\
\dot{\hat{\theta}}_3&=-\gamma_\theta (\hat{x}_0-x_0)\phi_{0,3}(x_0)
\end{split}
\end{equation}
\begin{equation}\label{eq:Morris_Lecar:nonlinear}
\begin{split}
\hat{\lambda}&= 3+ \hat{x}_{1,1}\\
\dot{\hat{x}}_{1,1}&=\gamma e \left(\hat{x}_{1,1} - \hat{x}_{2,1} - \hat{x}_{1,1}\left(\hat{x}_{1,1}^2+\hat{x}_{2,1}^2\right)\right)\\
\dot{\hat{x}}_{2,1}&=\gamma e \left(\hat{x}_{1,1}+ \hat{x}_{2,1} - \hat{x}_{2,1}\left(\hat{x}_{1,1}^2+\hat{x}_{2,1}^2\right) \right)\\
e&=\sigma(\|x_0-\hat{x}_0\|_\varepsilon)
\end{split}
\end{equation}
Parameters of the observer were set as follows: $\alpha=1$,
$\varepsilon=0.001$, $\gamma=0.01$, $\gamma_{\theta}=0.05$.

\begin{figure}
\centering
\includegraphics[width=400pt]{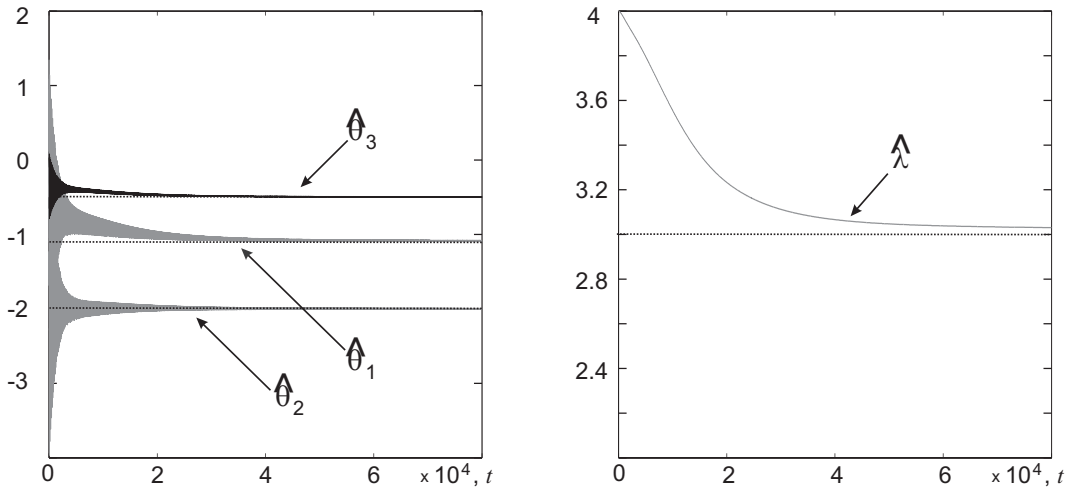}
\caption{Trajectories of the estimates of parameters $\theta_1$,
$\theta_2$, $\theta_3$ and $\lambda$ as functions of time.
 The periodic time of the spikes is circa $10$ seconds while the total
time simulated is $8\times 10^4$ seconds, thus the figure spans
thousands of spike cycles.
 The visual effect of these extreme time scales is seen in the figure;
 hunting oscillations within the observer are seen
as thickening of the graphs of certain estimated parameters. }
 \label{fig:Morris_Leacar}
\end{figure}

According to Theorems \ref{theorem:neural_identification},
\ref{theorem:parameter_identification} observer
(\ref{eq:Morris_Lecar:linear}),
(\ref{eq:Morris_Lecar:nonlinear}) should ensure successful
reconstruction of the model parameters provided that the
regressor is persistently exciting. This requirement is
satisfied for model (\ref{eq:Morris_Lecar:1}) generating
periodic solutions.  We simulated system
(\ref{eq:Morris_Lecar:1}),  (\ref{eq:Morris_Lecar:linear}),
(\ref{eq:Morris_Lecar:nonlinear}) over a wide range of initial
conditions. Figure \ref{fig:Morris_Leacar} shows an example of
typical behavior of the observer over time. As we can see from
this figure all estimates converge to small neighborhoods of
true values of the parameters.

\vspace*{0.5cm}
\section{Conclusion}\label{sec:conclude}

In this article we have reviewed and explored observer-based
approaches
 to the problem of state and parameter reconstruction for
 classes of typical models of neural oscillators.
 The estimation procedure in this approach is defined as a
 system of ordinary differential equations of which
the right-hand side does not depend explicitly on the
unmeasured variables. The solution of this system (or functions
of the solutions)
 should  asymptotically converge
 to small neighbourhoods of the actual values of the variables to be estimated.
 Until recently, due to nonlinear dependence
of the vector-fields of the models on unknown
 parameters and also due to uncertainties in the time scales of hidden variables,
 observer-based approach to solving the problem of state and parameter estimation
of neural oscillators was a relatively unexplored territory.
Here we demonstrate that despite these obvious difficulties the
approach can be successfully applied to a wide range of models.

Two different strategies to observer design have been studied
in the paper. The first strategy is based on the availability
of canonical representations
 of the original system. Success of this strategy is obviously determined by
 wether one can find a suitable coordinate transformation such that the
equations of the original model can be transformed into the
canonical adaptive
 observer form. Because a coordinate transformation is required, different
 classes of models are likely to lead to different observers. The second
strategy is based on the ideas and approaches of universal
adaptive regulation \cite{Dyn_Con:Ilchman:97}, non-uniform
convergence and non-uniform small-gain theorems
\cite{SIAM_non_uniform_attractivity},
 \cite{Adaptive_Observers}. The structure of observers obtained as a
 result of this design strategy does not change much from one model
to another. The main difference between these design strategies
is in the convergence rates: exponential for the first and
asymptotical
 for the second. As long as mere overall convergence time is accounted
for there is no big difference whether the convergence itself
is exponential or not. Yet, the fact that it can be made
exponential with known rates of convergence allows us to derive
the a-priori estimates of the amount of time needed to achieve
a certain given accuracy of estimation.

We have shown that for linearly parameterized models such as
the FitzHugh-Nagumo
 and Hindmarsh-Rose oscillators one can develop an observer for state and
parameter estimation of which the convergence rate is
exponential. For the
 nonlinearly parameterized and more realistic models such as the Morris-Lecar
and Hodgkin-Huxley equations we presented an observer of which
the convergence is asymptotic. In both cases the rate of
convergence depends on the degree of
 excitation in the measured data. In the case of linearly parameterized
systems this excitation can be measured by the minimal
eigenvalue of a
 certain matrix constructed explicitly from the data and the model.
For the nonlinearly parameterized systems the degree of
excitation is
 defined by a more complex expression, (\ref{eq:nonlinear_pe}).
In principle, one can ensure arbitrarily fast convergence of
the estimator
 provided that the excitation is sufficiently high. This property motivates
 the development of measurements protocols that are most consistent with a range
 of models that will be fitted to the collected data. In fact, in order to achieve
 higher computational effectiveness, one shall aim to produce data of which
 the excitation is higher for the given range of models.

One question remains unexplored though -- the actual amount of
elementary computational operations required to realize these
two observer schemes. This number depends substantially on the
required accuracy of estimation.
 We aim to answer this important question in future case studies.


\section{Appendix}\label{appendix}

{\it Proof of Theorem \ref{theorem:parameter:independent}}
\begin{proof} The proof is straightforward. Indeed, for the observability test we have
\begin{eqnarray}
 \left(
 \begin{array}{c}
  h(x) \\ L_f h(x) \\ L^2_f h(x) \\ . \\ . \\ . \\ L^{n-1}_f h(x)
 \end{array}
 \right)
 =
 \left(
  \begin{array}{c}
   x_1
  \\
   x_2+x_3+...+x_n
  \\
   0
  \\
   .
  \\
   0
  \end{array}
  \right)
\end{eqnarray}
and
\begin{eqnarray}
 \frac{\partial}{\partial x}
 \left(
  \begin{array}{c}
   h(x) \\ L_f h(x) \\ L^2_f h(x) \\ . \\ . \\ . \\ L^{n-1}_f h(x)
  \end{array}
 \right)
 =
 \left(
  \begin{array}{ccccccc}
   1 & 0 & 0 & 0 & 0 & ... & 0
  \\
   0 & 1 & 1 & 1 & 1 & ... & 1
  \\
   0 & 0 & 0 & 0 & 0 & ... & 0
  \\
   . & . & . & . & . & ... & .
  \\
   0 & 0 & 0 & 0 & 0 & ... & 0
  \end{array}
  \right)
\end{eqnarray}
It follows that observability is lost for $n > 2$. This proves
condition (i). In order to demonstrate condition (ii) we use
theorem \ref{Marino_Thm} as follows. From equation
(\ref{vector_g}) we have
\begin{eqnarray}
 \left\langle
 \left(
 \begin{array}{cc} 1 & 0 \\ 0 & 1 \end{array}
 \right),
 \left( \begin{array}{c} g_1\\g_2 \end{array} \right)
 \right\rangle
 =
 \left( \begin{array}{c} 0\\1 \end{array} \right)
\end{eqnarray}
giving
\begin{eqnarray}
 g =
 \left(
  \begin{array}{c} 0 \\ 1 \end{array}
 \right)
\end{eqnarray}
Theorem \ref{Marino_Thm} condition (i) will be satisfied since
the system is linear and from theorem \ref{Marino_Thm} part
(ii) we have
\begin{eqnarray}
 [q_1,ad^0_f g] &=& [q_1,g]
\\
 &=& \frac{\partial g}{\partial x}q_1
  - \frac{\partial q_1}{\partial x}g
\\
 &=&
 -\frac{\partial q_1}{\partial x}
 \left(
  \begin{array}{c} 0 \\ 1 \end{array}
 \right)
\\
 &=& -\frac{\partial q_1}{\partial x_2}
\end{eqnarray}
we satisfy theorem \ref{Marino_Thm} part (ii) if and only
 if $\partial{q_1} / {\partial x_2} = \mathbf{0}$ for all $x \in U$. 
\end{proof}

{\it Proof of Corollary \ref{cor:BG_HR_observer}}
\begin{proof}
The proof of the corollary is straightforward. Consider the
error system given by (\ref{eqn:BG_err_sys1}). Given  that the
function $\varphi(t)$ is bounded one can easily see that
$\hat{\eta}$, $\hat{x}$ are bounded as well (consider e.g. the
following Lyapunov candidate: $V=\|\tilde{x}^{\ast}\|^2 +
\tilde{\eta} \Gamma^{-1}\tilde{\eta}$). This implies that
component $\varphi_4(t)\tilde{\eta}_4(t)$ is converging to zero
exponentially.

Let us denote
\[
\bar{\eta}=(\tilde{\eta}_1,\tilde{\eta}_2,\tilde{\eta}_3,\tilde{\eta}_5,\tilde{\eta}_6,\tilde{\eta}_7,\tilde{\eta}_8)^{T}
\]
and consider the following reduced error dynamics
\begin{eqnarray}\label{eqn:BG_err_sys_reduced}
\begin{array}{c}
 \dot{\tilde{x}}^*
 =
 \left(
  \begin{array}{cc}
   -c_1 & k
\\
   0 & f
  \end{array}
 \right)
 \tilde{x}^* +
 \left(
  \begin{array}{c}
   \bar{\varphi}^T\bar{\eta}
\\
   0
  \end{array}
  \right) + p \varepsilon(t)
\\
 \dot{\bar{\eta}} = -\Gamma\bar{\varphi}\tilde{x}_1^*
\end{array}
\end{eqnarray}
in which $\varepsilon(t)$ stands for the term
$\varphi_4(t)\tilde{\eta}_4(t)$, and $p=(1,0)^{T}$. System
(\ref{eqn:BG_err_sys_reduced}) is a linear time-varying system
of which the homogenous part is exponentially stable provided
that $\bar{\varphi}(t)$ is persistently exciting (this follows
explicitly from Theorem \ref{theorem:BG}). Hence, taking into
account that $\varepsilon(t)$ is an exponentially converging to
zero term, we can conclude that $\tilde{x}^{\ast}$,
$\bar{\eta}$ converge to the origin too and that such
convergence is exponential.
\end{proof}

{\it Proof of Theorem \ref{theorem:observer:passivity}}
\begin{proof} The proof of the theorem is standard and can be constructed from many other more general results (see for example \cite{Marino90}, \cite{Narendra89}).
 Here we present just a sketch of the argument for consistency. According to our assumptions, matrix $A_1 + LC_1$ is Hurwitz. Moreover, the transfer function $$H(s)=C_1(s-(A_1 + LC_1))^{-1} b=\frac{s+k}{s^2 + (k+1) s + k}=\frac{s+k}{(s + k)(s+1)}$$
is strictly positive real. Hence, using the Kalman-Yakubovich-Popov lemma, we can conclude that there exists a symmetric and positive definite matrix $H$ such that
\begin{equation}\label{eq:proof:1}
H (A_1 + LC_1) + (A_1 + LC_1)^{T} H < - Q, \ H b = (1, 0)^{T},
\end{equation}
where $Q$ is a positive definite matrix. Let us now consider
the following function
\[
V(z,\hat{\upsilon},t)= \frac{1}{2}(z-\hat{z})^{T}H (z-\hat{z}) + \frac{1}{2}\|\upsilon-\hat{\upsilon}\|^2 \gamma^{-1} +  D\int_{t}^{\infty} \varepsilon^2(\tau)d\tau
\]
where the value of $D$ is to be specified later. Clearly, the
function $V$ is well-defined for the term $\varepsilon(t)$ is
continuous and exponentially decaying to zero as
$t\rightarrow\infty$. Thus the boundedness of $V$ implies that
$\|\hat{z}-z\|$ and $\|\hat{\upsilon}-\upsilon\|$ are  bounded.

Consider the time-derivative of $V$:
\[
\begin{split}
\dot{V} & = (z-\hat{z})^{T}(H(A_1+LC_1)+(A_1+LC_1)^{T}H)(z-\hat{z})+ (z-\hat{z})^{T} H b \phi^{T}(z_1,t)(\upsilon-\hat{\upsilon})\\&+ (z-\hat{z})^{T} H b \varepsilon
 - (\upsilon-\hat{\upsilon}) (1,0) (z-\hat{z}) \phi(z_1,t) - D \varepsilon^2
\end{split}
\]
Taking (\ref{eq:proof:1}) into account we obtain:
\begin{equation}\label{eq:proof:2}
\begin{split}
\dot{V}&\leq - (z-\hat{z})^T Q  (z-\hat{z}) + \|z-\hat{z}\| |\varepsilon(t)| M - D \varepsilon^2\leq -\alpha\|z-\hat{z}\|^2 +\\
& \|z-\hat{z}\| |\varepsilon(t)| M - D \varepsilon^2,
\end{split}
\end{equation}
where $\alpha>0$ is the minimal eigenvalue of $Q$,  $M$  is
 fixed positive number, and $D$ is a parameter of $V$ which can
 be chosen arbitrarily and independently of $M$. Choosing $D$ such that
\[
\sqrt{\frac{\alpha D}{4}}=M,
\]
we ensure that
\[
\dot{V}\leq -\frac{\alpha}{2}\|z-\hat{z}\|^2
\]
Thus the function $V$ is bounded from above, and given that the
solution $z(t)$ exists for all $t>t_0$ so does the solution of
the combined system. The rest of the proof follows directly
from Barbalatt's lemma and the asymptotic stability theorem for
the class of skew-symmetric time-varying systems presented in
\cite{Morgan_77}.
\end{proof}

\end{document}